\documentclass[aps,prc,preprint,floatfix,showpacs,preprintnumbers,amsmath,amssymb]{revtex4-1}
\usepackage{graphicx}
\usepackage{dcolumn}
\usepackage{bm}
\usepackage{xcolor}
\usepackage{graphicx}
\usepackage{color}
\usepackage[outdir=./]{epstopdf}
\usepackage[toc,page]{appendix}
\usepackage{multirow}

\newcommand{\lp}{\left}
\newcommand{\rp}{\right}

\begin{document}

\title{Nuclear charge-exchange excitations based on relativistic density-dependent point-coupling model}
\author{D. Vale}
\email{deni.vale1@skole.hr}
\affiliation{Gimnazija i strukovna \v skola Jurja Dobrile Pazin, \v Setali\v ste Pazinske gimnazije 11, Pazin 52000, Croatia}
\affiliation{Osnovna \v skola Vodnjan/Scuola elementare Dignano, \v Zuka 6/Via delle ginestre 6, Vodnjan/Dignano 52215, Croatia}
\affiliation{Department of Physics, Faculty of Science, University of Zagreb, Bijeni\v cka cesta 32, 10000 Zagreb, Croatia}
\author{Y. F. Niu}
\affiliation{School of Nuclear Science and Technology, Lanzhou University, Lanzhou 730000, China, \\ ELI-NP, Horia Hulubei National Institute for Physics and Nuclear Engineering, 30 Reactorului Street, RO-077125, Bucharest-Magurele, Romania}
\author{N. Paar}
\email{npaar@phy.hr}
\affiliation{Department of Physics, Faculty of Science, University of Zagreb, Bijeni\v cka cesta 32, 10000 Zagreb, Croatia}
\date{\today}

\begin{abstract}
Spin-isospin transitions in nuclei away from the valley of stability are essential 
for the description of astrophysically relevant weak interaction processes. While they remain mainly beyond the reach of 
experiment, theoretical modeling provides important insight into their properties. In order to describe
the spin-isospin response,
the proton-neutron relativistic quasiparticle random phase approximation (PN-RQRPA)
is formulated using the relativistic density-dependent point coupling interaction, and separable pairing interaction
in both the $T=1$ and $T=0$ pairing channels. By implementing recently established DD-PCX interaction with improved
isovector properties relevant for the description of nuclei with neutron-to-proton number asymmetry, 
the isobaric analog resonances (IAR) and Gamow-Teller resonances (GTR) have been investigated. In contrast to other models that usually underestimate the IAR excitation energies in Sn isotope chain, the present
model accurately reproduces the experimental data, while the GTR properties depend on the isoscalar pairing interaction strength. 
This framework provides not only an improved description of the spin-isospin response in nuclei,  but it also allows 
future large scale calculations of charge-exchange excitations and weak interaction processes in stellar environment.
\end{abstract}

\maketitle

\bigskip \bigskip
%=========================================================================
%  Section 1

\section{\label{secI}Introduction}

Charge-exchange excitations in atomic nuclei correspond to a class of nuclear transitions
composed of the particle-hole configurations that contain the exchange of the nucleon 
charge, 
described by the isospin projection lowering (increasing) operator $\tau_-$ ($\tau_+$).
The fundamental charge-exchange excitation is the isobaric analog resonance (IAR)~\cite{Auerbach1972,Auerbach1983,Rumyantsev1994,Pham1995,Colo1998,Paar2004,fracasso2005,roca-maza12b}, 
with no changes in quantum numbers $\Delta J = \Delta L = \Delta S =\text{0}$, thus the IAR corresponds to a collective 
excitation with $J^{\pi}={\text{0}}^+$.
The Gamow-Teller resonance (GTR) represents another relevant charge-exchange mode, characterized by $J^{\pi}={\text{1}}^+$, i.e., it 
corresponds to spin-flip excitations without changing the orbital motion, $ \Delta S =\text{1}$,  $ \Delta L =\text{0}$.

As it has been emphasized in Ref.~\cite{XaviNils.2018}, recent interest in the GTR studies is motivated by its importance for
understanding the spin and spin-isospin dependence of modern effective interactions 
~\cite{Bender2002,Marketin2012a,Litvinova2014,Ekstrom2014,Zelevinsky2017,Morita2017,Nabi2017,Ha2017}, nuclear beta decay~\cite{Eng.88,Bor.95,Eng.99,Borzov2003,Madruga2016,Marketin2016,Moon2017,Wang_2016,Ravlic2020}, beta delayed neutron
emission~\cite{Folch2016}, as well as double beta decay~\cite{Faessler1998,Suhonen1998,Simkovic2011,Menendez2011,Vergados2012,Menendez2014,Stefanik2015,Navas2015,Suhonen2017}. In addition, accurate description of GT$^{\pm}$ transitions, including both in stable and exotic nuclei, is relevant for the description of a variety of  astrophysically relevant weak interaction processes~\cite{Langanke2000,Langanke2003,Janka2007,Noji2014,Paar2015}, 
electron capture in presupernova stars ~\cite{Niu2011,Niu2013,Ravlic_ec2020},
r-process~\cite{Arnould2007,Mori2016} and neutrino-nucleus interaction of relevance for neutrino detectors and neutrino nucleosynthesis in stellar environment~\cite{Ejiri2000,Suzuki2003,Frekers2011,Cheoun2010,Paar2008,Paar2011,Karakoc2014}.

The properties of charge-exchange modes of excitation have extensively been studied ~\cite{Osterfeld1992,Fujita2011}. %
Following theoretical prediction ~\cite{Ikeda1963},  in 1975 the GTR has been experimentally confirmed in $(p,n)$ 
reactions~\cite{Doering1975}. The  GTR represents one of the most extensively investigated collective excitation 
in nuclear physics, both experimentally and theoretically (e.g., see Refs.~\cite{Osterfeld1992,Halbleib1967,Doering1975,Towner1979,Goodman1980,Brown1981,Bertsch1982,Nakayama1982,Gia.81,Gaarde1983,Kuzmin1984,Col.94,Ham.93,Brown1994,Suzuki1997,Conti1998,Caurier1999,Langanke2001,Bender2002,Algora2003,Kalmykov2006,Bai2009,Lutostansky2011,Sasano2011,Fujita2011,Niu2012,Ha2013,roca-maza13b,Martini2014,Ha2016,Niu2016,Liang2018,Yasuda2018}). More details about 
experimental studies of spin-isospin excitations are also reviewed in Ref. \cite{Fujita2011}. Recent studies of the GTR in the framework based on relativistic energy density functional include relativistic quasiparticle random phase approximation (RQRPA)~\cite{Paar2004},  the relativistic RPA based on the relativistic Hartree-Fock (RHF) \cite{Liang2008},  
relativistic QRPA based on point coupling model with nonlinear interactions \cite{Niu2013,Wang_2016},
and relativistic QRPA formulated using the relativistic Hatree–Fock–Bogoliubov (RHFB) model for the ground state \cite{Niu2017}. In Ref.~\cite{Finelli2007} the nuclear density functional framework, based on chiral dynamics and the symmetry breaking pattern of low-energy QCD, has been used to formulate the proton–neutron QRPA to investigate the role of chiral pion–nucleon dynamics in the description of charge-exchange excitations. 
The GTR has also been studied by including couplings between single nucleon and collective nuclear vibrations, e.g., particle-vibration coupling (PVC) 
of the 1p-1h$\otimes$phonon type of coupling \cite{Niu2012,Marketin2012,Niu2016,Rob2019}. The PVC allows to include important dynamical correlations 
missing in the static self-consistent mean field models and it provides additional fragmentation of the GTR strength when compared to the
random phase approximation studies based only on 1p-1h configurations \cite{Niu2012}.

At present, the knowledge about charge-exchange transitions in nuclei away from the valley of stability is 
rather limited, and mainly beyond the reach of experiment.  Since these nuclei are especially important for
their astrophysical relevance in stellar evolution and
nucleosynthesis, it is crucial to develop microscopic theoretical approaches, that allow
quantitative and systematic analyses of the transition strength distributions of 
unstable nuclei. In order to assess the overview into systematical model uncertainties 
in modeling charge-exchange excitation phenomena, it is important to address 
their properties from various approaches, by implementing different theory frameworks and
effective nuclear interactions.

In Ref.~\cite{Paar2004} charge-exchange excitations have been 
studied in the framework based on the relativistic nuclear energy density functional (EDF),
within the approach that unifies the treatment of mean-field and
pairing correlations, relativistic quasiparticle random phase approximation (RQRPA)
formulated in the canonical single-nucleon basis of the relativistic 
Hartree-Bogoliubov (RHB) model. 
In this implementation the relativistic EDF with explicit density dependence of the meson-nucleon
couplings is used, that provides an improved description of asymmetric nuclear matter, 
neutron matter and nuclei far from stability. The pairing correlations were 
described by the pairing part of the finite range Gogny interaction \cite{GognyD1S,Berger1991}.  
However, the EDFs have usually been parameterized with the experimental 
data on the ground state properties, supplemented with the pseudo-observables
on nuclear matter properties. In the case of density dependent meson-exchange
interactions, the neutron skin thickness in $^{208}$Pb has been introduced as
an additional constraint on the isovector channel of the effective interaction.
However, the results from measurements of the neutron-skin thickness are usually
model-dependent, and the pseudo-observables on nuclear matter are often
rather arbitrary. Recently, a novel EDF parameterization has been established 
based on the relativistic point coupling interaction, by using in the $\chi^2$ minimization
the nuclear ground state properties (binding energies, charge radii, pairing gaps)
together with the properties of collective excitations in nuclei, isoscalar giant monopole
resonance energy and dipole polarizability~\cite{Yuksel.09}. In this way an 
effective interaction DD-PCX has been established with improved isovector
properties, that is successful not only in the description of nuclear ground state,
but also of the excitation phenomena, incompressibility of nuclear matter and the
symmetry energy close to the saturation density~\cite{Yuksel.09}. The improved isovector channel
for the DD-PCX interaction is especially important not only for the symmetry energy
of the nuclear equation of state, but also for the description of ground-state and
excitation properties of $N\neq Z$ nuclei. Clearly, this is very important for the 
implementation of the EDF based models to exotic nuclei, as well as for applications
in nuclear astrophysics.

In this work we establish the proton-neutron RQRPA in the canonical single-nucleon basis 
of the RHB model based on density-dependent relativistic point coupling
interaction. 
Our study represents the first implementation of the relativistic point 
coupling interaction with density dependent vertex functions in formulating the RQRPA
for the description of charge-exchange excitations.
In addition, the treatment of the pairing correlations is also improved, 
by implementing the separable pairing force that allows accurate and efficient calculations 
of the pairing properties~\cite{Tian2009}.  By using recently established DD-PCX
interaction with improved properties that are essential for description of nuclei away from 
the valley of stability~\cite{Yuksel.09}, the proton-neutron RQRPA established in this work
will be employed in the investigation of the properties of collective charge-exchange excitations,
IAR and GTR.

Clearly, a study of both the IAR and GTR properties represents an important benchmark test
for novel theoretical approaches established not only for description of charge-excitation modes,
but also for modeling a variety of astrophysically relevant processes in stellar environment.
Therefore, in the present work that introduces a microscopic approach to describe 
charge-exchange excitations based on density-dependent relativistic point coupling interaction, the novel theory framework will be 
employed in the analyses of charge-exchange modes, the IAR and GTR, both for magic nuclei, 
as well as for open-shell nuclei to probe the effect of the pairing correlations.

In Sec.~\ref{secII} the formalism of the proton-neutron RQRPA based on
the density dependent point coupling interaction is introduced. In Sec.~\ref{secIII} the model is employed
in studies of charge-exchange modes of excitation, the IAR and the GTR. 
The conclusions of this work are summarized in Section~\ref{secIV}.

%%%%%%%%%%%%%%%%%%%%%%%%%%%%%%%%%%%%%%%%%%%%%%%%%%%%%%%%%%%%%%%%%%%%%%%%%%
%=========================================================================
%  Section 2

\section{\label{secII}Proton-neutron RQRPA based on relativistic point coupling interaction}
%=========================================================================

In the previous implementation in the relativistic framework, the RQRPA has
been established on the ground of relativistic 
Hartree-Bogoliubov (RHB) model,
based on the effective Lagrangian with density-dependent meson-exchange
interaction terms~\cite{Paar2004}. Therein the pairing correlations have been
described by the pairing part of the Gogny interaction~\cite{GognyD1S,Berger1991}.
In the present study, the RQRPA is established using the
relativistic point coupling interaction, while the pairing correlations 
are described by the separable pairing force from Ref.~\cite{Tian2009}. 
Since the full RQRPA equations are rather complicated, in the present
study we solve the respective equations in the canonical basis,
where the Hartree-Bogoliubov wave functions can be expressed in the
form of the BCS-like wave functions. More details 
on the implementation of the canonical basis in the RHB model and
general formalism of the PN-RQRPA equations in the canonical basis 
are given in Ref.~\cite{Paar2004}. The focus of this work is the
implementation of the relativistic point coupling interaction in deriving
the PN-RQRPA equations. The nuclear ground state properties are
described in the RHB model for the point coupling interaction, described
in detail in Ref.~\cite{Nik.14}.

Starting from the $0^+$ ground state of a spherical even-even nucleus,
transitions to $J^\pi$ excited state of the corresponding odd-odd 
daughter nucleus are considered, using the charge-exchange operator
${\cal{O}}^{JM}$. The general form of the PN-RQRPA equations 
read~\cite{Paar2004},

\begin{equation} \left( \begin{array} [c]{cc}
A^{J} & B^{J}\\
B^{^{\ast}J} & A^{^{\ast}J}
\end{array}
\right)  \left( \begin{array} [c]{c}
X^{\lambda J}\\
Y^{\lambda J}
\end{array}
\right)  =E_{\lambda}\left( \begin{array} [c]{cc}
1 & 0\\
0 & -1
\end{array}
\right)  \left( \begin{array} [c]{c}
X^{\lambda J}\\
Y^{\lambda J}
\end{array}\right) \; ,
\label{pnrqrpaeq}
\end{equation}
where the  $A$ and $B$ matrices are defined in the canonical basis,
\begin{eqnarray}
A_{pn,p^\prime n^\prime}^{J} &=& H^{11}_{pp^\prime}\delta_{nn^\prime} +
  H^{11}_{nn^\prime}\delta_{pp^\prime}  \nonumber \\ & & +
\lp( u_p v_n u_{p^\prime} v_{n^\prime} + v_p u_n v_{p^\prime} u_{n^\prime}\rp)
 V_{pn^\prime n p^\prime}^{ph J} + 
\lp( u_p u_n u_{p^\prime} u_{n^\prime} + v_p v_n v_{p^\prime} v_{n^\prime}\rp) 
 V_{pn p^\prime n^\prime}^{pp J} \nonumber \\
B_{pn,p^\prime n^\prime}^{J} &=& 
\lp( u_p v_n v_{p^\prime} u_{n^\prime} + v_p u_n u_{p^\prime} v_{n^\prime}\rp)
 V_{p n^\prime n p^\prime}^{ph J} \nonumber \\ & &- 
\lp( u_p u_n v_{p^\prime} v_{n^\prime} + v_p v_n u_{p^\prime} u_{n^\prime}\rp) 
 V_{pn p^\prime n^\prime}^{pp J} \; .
\label{abmat}
\end{eqnarray}
The proton and neutron quasiparticle canonical states are denoted by $p$, $p\prime$, and $n$, $n\prime$, respectively.
$V^{ph}$ is the proton-neutron particle-hole residual
interaction, and $V^{pp}$ is the corresponding particle-particle interaction, and
$u$ and $v$ denote the occupation amplitudes of the respective states.
Since the canonical basis does not diagonalize the Dirac single-nucleon
mean-field Hamiltonian, the off-diagonal matrix elements 
$H^{11}_{nn^\prime}$ and $H^{11}_{pp^\prime}$ are also included in
the $A$ matrix, as given in Ref.~\cite{Paar2004}.
$E_{\lambda}$ denote the excitation energy, while $X^{\lambda J}$ and $Y^{\lambda J}$
are the corresponding forward- and backward-going QRPA amplitudes,
respectively. 

By solving the eigenvalue problem (\ref{pnrqrpaeq}), the reduced transition strength
can be obtained between the ground state of the even-even
$(N,Z)$ nucleus and the excited state of the odd-odd $(N+1,Z-1)$ or $(N-1,Z+1)$ 
nucleus, using the corresponding transition operators ${\cal{O}}^{JM}$ in both channels,
\begin{align}
B_{\lambda J}^{-} = \lp| \sum_{pn} <p||{\cal{O}}^J||n> 
\lp( X_{pn}^{\lambda J} u_p v_n + Y_{pn}^{\lambda J}v_p u_n \rp) \rp|^2
\; ,
\label{strength-}\\
B_{\lambda J}^{+} = \lp| \sum_{pn} (-1)^{j_p+j_n+J}<n||{\cal{O}}^J||p> 
\lp( X_{np}^{\lambda J} v_p u_n +  Y_{np}^{\lambda J}u_p v_n \rp) \rp|^2
\; .
\label{strength+}
\end{align}
For the presentation purposes, the discrete strength distribution is folded by the 
Lorentzian function of the width $\Gamma=1$ MeV,
\begin{equation}
R(E)^{\pm} = \sum_{\lambda}B_{\lambda J}^{\pm}\frac{1}{\pi}
\frac{\Gamma/2}{(E-E_{\lambda_{\pm}})^2+(\Gamma /2)^2} \; .
\label{lorentzian}
\end{equation}

In the implementation of the relativistic point coupling interaction, 
the spin-isospin dependent terms in the residual interaction of the
PN-RQRPA are induced by the isovector-vector and pseudovector
terms. In comparison, for the finite range meson-exchange interaction,
these terms were obtained from the $\rho$-and $\pi$-meson exchange,
respectively~\cite{Paar2004}. In the present study, the PN-RQRPA 
residual interaction terms $V_{abcd}$ are derived from the effective 
Lagrangian density for the point coupling interaction.

The isovector-vector part of PN-RQRPA contains only non-rearrangement terms of the residual two-body interaction. 
For the respective spacelike components we obtain,
\begin{equation}
V_{abcd}^{(\text{TVs})}~=~-\int d^3 r_1 \int d^3 r_2 \psi_a^\dagger (\vec{r}_1)\left(\vec{\tau}\gamma_0\gamma_i\right)^{(1)}\psi_c(\vec{r}_1)\alpha_{TV}(\rho)\delta(\vec{r}_1-\vec{r}_2)\psi_b^\dagger (\vec{r}_2)\left(\vec{\tau}\gamma_0\gamma^i\right)^{(2)}\psi_d(\vec{r}_2),
\end{equation}
and timelike components are given by
\begin{equation}
V_{abcd}^{(\text{TVt})}~=~\int d^3 r_1 \int d^3 r_2 \psi_a^\dagger (\vec{r}_1)\vec{\tau}^{(1)}\psi_c(\vec{r}_1)\alpha_{TV}(\rho)\delta(\vec{r}_1-\vec{r}_2)\psi_b^\dagger (\vec{r}_2)\vec{\tau}^{(2)}\psi_d(\vec{r}_2),
\end{equation}
where the coupling $\alpha_{TV}(\rho)$ is a function of baryon (vector) density.
For Dirac spinor given by,
\begin{equation}
    \psi_{a m_a}(r) ~=~ \left(\begin{array}{clcr}
f_a(r) \Omega_{\kappa_a m_a}(\Omega) \\\
ig_a(r) \Omega_{\bar{\kappa}_a m_a}(\Omega) \end{array}\right),
\end{equation}
where $a$ stands for all quantum numbers, with the exception of the projection of total angular momentum $m_a$. Quantum number $\kappa$ is defined as $\kappa=-(l+1)$ for $j = l + 1/2$ and $\kappa=l$ for $j = l -1/2$, while $\bar{l} = 2j - l$ corresponds to the lower component orbital angular momentum \cite{Gre.00}.
The spacelike part of the matrix elements obtained by the angular momentum coupling is given
\begin{align}
&V_{abcd}^{(\text{TVs})J}~=~\frac{2}{ \left(2 J +1 \right)}\sum_L \int dr r^2 \alpha_{TV}(\rho) 
\bigg[f_a(r)g_c(r)\langle (1/2\ l_a) j_a || \left[\sigma_S Y_L\right]_J || (1/2\ \bar{l}_c) j_c \rangle - \nonumber\\
&~g_a(r)f_c(r)\langle (1/2\ \bar{l}_a) j_a || \left[\sigma_S Y_L\right]_J || (1/2\ l_c) j_c \rangle \bigg]  \bigg[f_b(r)g_d(r)\langle (1/2\ \bar{l}_d) j_d || \left[ \sigma_S Y_L\right]_J || (1/2\ l_b) j_b \rangle\nonumber\\ &~- g_b(r)f_d(r)\langle (1/2\ l_d) j_d || \left[\sigma_S Y_L\right]_J || (1/2\ \bar{l}_b) j_b \rangle \bigg],
\label{spacelike}
\end{align}
and the timelike part is
\begin{align}
V_{abcd}^{(TVt)J}~=&~\frac{2}{2J + 1} \int dr r^2 \alpha_{TV}(\rho)\bigg[f_a(r)f_c(r) + g_a(r)g_c(r)\bigg]\bigg[f_b(r) f_d(r) + g_b(r) g_d(r)\bigg] \nonumber \\ &~\times \langle (1/2\ l_a) j_a || Y_J || (1/2\ l_c) j_c \rangle \langle (1/2\ l_d) j_d || Y_J || (1/2\ l_b) j_b \rangle. 
\label{timelike}
\end{align}
When compared to standard R(Q)RPA matrix elements, in particular corresponding direct term, there exists an additional factor of 2 in the numerator of Eqs. \eqref{spacelike} and \eqref{timelike} due to the difference in the isospin part of the matrix element (see Appendix \ref{phelements}).
The isovector-pseudovector part of the point coupling interaction is given by 
\begin{equation}
V_{PV}~=~-\alpha_{PV}\delta(\vec{r}_1-\vec{r}_2) \left(\gamma_0\gamma_5\gamma_\mu\vec{\tau}\right)^{(1)}
\left(\gamma_0\gamma_5\gamma^\mu\vec{\tau}\right)^{(2)},
\end{equation}
where $\alpha_{PV}$ denotes the strength parameter of the interaction.
Since $\alpha_{PV}$ remains a free parameter of the model that cannot be constrained by the ground state properties,
its value should be constrained by the experimental data on charge-exchange excitations.
For the corresponding timelike part of the pseudovector matrix elements we obtain,
\begin{align}
V_{abcd}^{\text{PV(t)J}}~=&~\frac{2 \alpha_{PV}}{2 J + 1}
\int dr r^2 \bigg[f_a(r)g_c(r) - g_a(r)f_c(r)\bigg] \bigg[f_b(r)g_d(r) - g_b(r)f_d(r)\bigg]\nonumber \\ &\times\langle (1/2\ l_a) j_a || Y_J || (1/2\ \bar{l}_c) j_c\rangle
\langle (1/2\ \bar{l}_d) j_d || Y_J || (1/2\ l_c) j_c\rangle.
\end{align}
We note that the timelike matrix elements are non-zero only in the case of unnatural parity transitions, e.g., for Gamow-Teller transitions. The spacelike part of the isovector-pseudovector matrix elements results,
\begin{align}
&V_{abcd}^{PV(s)J}~=~\frac{2\alpha_{PV}}{2 J + 1} 
\sum_L\int dr r^2 \bigg[ f_a(r)f_c(r)\langle (1/2\ l_a) j_a ||\left[\sigma_S Y_L\right]_J|| (1/2\ l_c)j_c \rangle \nonumber\\ &~+~ g_a(r)g_c(r) \langle (1/2\ \bar{l}_a) j_a ||\left[\sigma_S Y_L\right]_J|| (1/2\ \bar{l}_c) j_c \rangle\bigg] \bigg[ f_b(r)f_d(r)\langle (1/2\ l_d) j_d || \left[\sigma_S Y_L\right]_J || (1/2\ l_b) j_b \rangle\nonumber\\ &~ +~ g_b(r)g_d(r)\langle (1/2\ \bar{l}_d) j_d || \left[\sigma_S Y_L\right]_J || (1/2\ \bar{l}_b) j_b \rangle\bigg] .
\end{align}
In order to constrain the value of the pseudovector coupling $\alpha_{PV}$, we follow the same procedure as used
in the case of relativistic functionals with meson-exchange \cite{Paar2004}, i.e., $\alpha_{PV}$ is adjusted to reproduce
the experimental value of excitation energy for Gamow-Teller resonance in $^{208}$Pb, $E=19.2$ MeV \cite{Horen1980,Aki.95, Kra.01}. In this way, we obtain  $\alpha_{PV}=$ 0.734 for DD-PC1, and $\alpha_{PV}=$ 0.621 
for DD-PCX point coupling interaction, and use these values systematically in all further investigations.

The PN-RQRPA also includes the pairing correlations. In most of the previous
applications of the RHB+RQRPA model, the pairing correlations have been 
described by the paring part of the Gogny force D1S~\cite{GognyD1S,Berger1991}.
This interaction has already been used in the RHB calculations of
various ground state properties in nuclei~\cite{Vretenar2005}.
Since the calculations based on the finite range Gogny force require 
considerable computational effort, in the present formulation of the
PN-RQRPA the separable form of the pairing interaction is used \cite{Tian2009}.
In Ref.~\cite{Tian2009} Y. Tian et al. introduced separable pairing interaction
in the gap equation of symmetric nuclear matter in $^{1}S_0$ channel:
\begin{equation}
\Delta(k)~=~-\int_0^\infty \frac{k'^2 dk'}{2\pi^2}\langle k | V_{sep}^{^{1}S_0} | k' \rangle \frac{\Delta(k')}{2 E(k')},
\end{equation}
where
\begin{equation}
\langle k | V_{sep}^{^{1}S_0} | k' \rangle ~=~ -G_0 p(k) p(k'),
\end{equation} 
with Gaussian ansatz
\begin{equation}
p(k) ~=~ e^{-a^2 k^2}.
\end{equation}
The two parameters $a$ and $G_0$ were adjusted to density dependence of the gap at the Fermi 
surface in nuclear matter, calculated with the Gogny force~\cite{Tian2009}. 
After transformation of the pairing force from momentum to coordinate space one obtains,
\begin{equation}
V(\vec{r}_1, \vec{r}_2, \vec{r}_1', \vec{r}_2')~=~-G_0 \delta(\vec{R}_1-\vec{R}_2) G(r) G(r') \frac{1-\hat{P}^{\sigma}}{2}
\label{pair_int}
\end{equation}
where $\vec{r} = 1/\sqrt{2}(\vec{r}_1 - \vec{r}_2)$ and $\vec{R} = 1/\sqrt{2}(\vec{r}_1 + \vec{r}_2)$. $G(r)$ is the 
Fourier transform of $p(k)$,
\begin{equation}
G(r)~=~\frac{e^{-r^2/(2a^2)}}{(4\pi a^2)^{3/2}}.
\label{Gaussian}
\end{equation}
Thus, the pairing force has finite range, and due to the presence of the factor $\delta(\vec{R}_1-\vec{R}_2)$ it
preserves the translational invariance~\cite{Tian2009}.
Due to coordinate transformation from laboratory to center of mass system and relative coordinates we need to use Talmi-Moschinsky brackets,
\begin{equation}
| n_1 l_1, n_2 l_2;\lambda \mu \rangle = \sum_{N L n l} M_{n_1 l_1 n_2 l_2}^{N L n l}|N L, n l; \lambda \mu \rangle.
\end{equation}
The definition of $M^{NLnl}_{n_1 l_1 n_2 l_2}$ is given in Refs. \cite{BUCK1996,KAM2001}. 
If the transformation matrix between two laboratory coordinates $\vec{r}_1$ and $\vec{r}_2$, and center of mass $\vec{R}$ and relative coordinate $\vec{r}$ is given
\begin{equation}
\left(\begin{array}{clcr}
\vec{R}\\
\vec{r}\end{array}\right)~=~
\left(\begin{array}{clcr}
\sqrt{\frac{d}{1+d}} & \sqrt{\frac{1}{1+d}}\\
\sqrt{\frac{1}{1+d}} & -\sqrt{\frac{d}{1+d}}\end{array}\right)
\left(\begin{array}{clcr}
\vec{r}_1\\
\vec{r}_2\end{array}\right),
\end{equation}
as a function of transformation parameter $d$ \cite{KAM2001}, then the most general definition of coefficient $M^{NLnl}_{n_1 l_1 n_2 l_2}$ is given by
\begin{align}
&M^{NLnl}_{n_1 l_1 n_2 l_2}(d)~=~i^{-(l_1+l_2+L+l)} 2^{-(l_1 + l_2 + L + l)/4} \sqrt{n_1 ! n_2 ! N ! n ![2(n_1+l_1)+1]!! [2(n_2+l_2)+1]!!}\nonumber\\
&~~~\times \sqrt{[2(N+L)+1]!! [2(n+l)+1]!!} \sum_{a b c d l_a l_b l_c l_d} (-1)^{l_a + l_b + l_c} (-1)^{(l_a + l_b + l_c + l_d)/2} d^{(2a+l_a+2d+l_d)/2}\nonumber\\
&~~~\times
 \frac{[(2l_a+1)(2l_b+1)(2l_c+1)(2l_d+1)]}{a! b! c! d! [2(a+l_a)+1]!![2(b+l_b)+1]!![(2(c+l_c)+1]!![2(d+l_d)+1]!!}\nonumber\\
 &~~~\times
(1+d)^{-(2a+l_a+2b+l_b+2c+l_c+2d+l_d)/2}\langle l_a 0 l_c 0 | L 0\rangle \langle l_b 0 l_d 0 | l 0 \rangle \langle l_a 0 l_b 0 | l_1 0 \rangle \langle l_c 0 l_d 0 | l_2 0 \rangle\nonumber\\
&~~~\times
\left\lbrace\begin{array}{clcr}
l_a & l_b & l_1\\
l_c & l_d & l_2\\
L & l & \Lambda
\end{array}\right\rbrace.
\end{align}
By employing the basis of spherical harmonic oscillator,
\begin{equation}
\tilde{I}_n ~=~\sqrt{4\pi} \int R_{nl}(r) G(r) r^2 dr~=~\frac{1}{2^{2/3}\pi^{3/4}b^{3/2}} \frac{(1-\alpha^2)^n}{(1+\alpha^2)^{n+3/2}} \frac{\sqrt{2n+1}}{2^n n!}, 
\end{equation}
the coupled matrix element for $T = 1$ pairing is given by,
\begin{align}
V_{abcd}^{(pair) JM}~=&~ -G_0\hat{j}_a\hat{j}_b\hat{j}_c\hat{j}_d (-1)^{l_b+l_d+j_a+j_c}
\left\{\begin{array}{clcr}
l_a & j_a & 1/2\\
j_b & l_b & J\end{array}\right\}
\left\{\begin{array}{clcr}
l_c & j_c & 1/2\\
j_d & l_d & J\end{array}\right\}\nonumber\\
&~\sum_{N n n'} \tilde{I}_n \tilde{I}_{n'} M_{n_a l_a n_b l_b}^{N J n 0} M_{n_c l_c n_d l_d}^{N J n' 0}.
\label{justT1pairing}
\end{align}
Furthermore, for Gamow-Teller transitions in open shell nuclei we need to extend Eq. \eqref{justT1pairing} to include both $T = 0$ and $T = 1$ channels. Therefore, we introduce natural extension of the pairing:
\begin{align}
V_{abcd}^{(pair)J M}~=&~- G_0\hat{j}_a \hat{j}_b \hat{j}_c \hat{j}_d \sum_{L S} \sum_{T}\frac{1}{2}\left(1+(-1)^{S'+T+1}\right) \tilde{f}(S,T) \hat{S}^2 \hat{L}^2\left\{\begin{array}{clcr}
l_b & 1/2 & j_b\\
l_a & 1/2 & j_a\\
L & S & J\end{array}\right\}\nonumber
\\
~&~\left\{\begin{array}{clcr}
l_d & 1/2 & j_d\\
l_c & 1/2 & j_c\\
L & S & J\end{array}\right\} \sum_{n n'} \tilde{I}_n \tilde{I}_n' M_{n_a l_a n_b l_b}^{N L n 0} M_{n_c l_c n_d l_d}^{N L n' 0}.
\label{total_pairing}
\end{align}
This is non-vanishing only for $S = 0$ and $T = 1$ or $S = 1$ and $T = 0$ pairing. Therefore, $\tilde f(S = 0, T = 1)~=~ 1$ case corresponds to the Eq. \eqref{justT1pairing}, while the case $\tilde f(S = 1, T = 0)~=~ V_{0pp}$. See Appendix  \ref{secpair} for detailed derivation. We don't know \textit{a priori} the value of the isoscalar proton-neutron pairing strength parameter $V_{0pp}$. It may be somewhat reduced or enhanced compared to the case $T = 1 (S = 0)$, which is only present at the RHB level, and should be deduced from experimental data on excitations or charge-exchange processes 
in open shell nuclei. 

In comparison to the nuclear ground state based on
the RHB, the PN-RQRPA residual interaction includes an additional
channel, described by the pseudovector term and additional $T = 0 ~(S = 1)$ pairing 
term for Gamow-Teller transitions in open shell nuclei, that are not present in
the ground state calculations. Apart from this, the PN-RQRPA introduced
in this work is self-consistent, i.e., the same interactions, both in the 
particle-hole and particle-particle channels, are used in the RHB 
equation that determines the canonical quasiparticle basis, and in 
the PN-RQRPA. In both channels, the same strength parameters 
of the interactions are used in the RHB and RQRPA calculations.  

Similar to the previous implementations of the PN-RQRPA~\cite{Paar2004},
the two-quasiparticle configuration space includes states with
both nucleons in the discrete bound levels, states with one nucleon in the
bound levels and one nucleon in the continuum, and also states with both
nucleons in the continuum.  The RQRPA configuration space also includes
pair-configurations formed from the fully or partially occupied
states of positive energy and the empty negative-energy states from the
Dirac sea~\cite{Paar2004}. As pointed out in Ref.~\cite{Paar2004}, the inclusion
of configurations built from occupied positive-energy states and empty 
negative-energy states is essential for the consistency of the model, as well
as to reproduce the model independent sum rules.

Model calculations are based on 20 oscillator shells in the RHB
model, as given as standard in Ref. \cite{Nik.14}
to achieve the convergence of the results.
At the level of the PN-RQRPA calculations no truncations in the maximal excitation energy for 
the $2qp$ configuration space are used. We included a truncation of the $2qp$ configuration 
space by the condition on the corresponding occupation factors \cite{Paar2004}, $u_{qp1}v_{qp2} > 0.001$,
in order to exclude $2qp$ configurations with two almost empty states. Further reducing of this
limit value does not modify the results.

%=========================================================================
%  Section 

\section{\label{secIII} Results}

\subsection{The isobaric analog resonance}

\begin{figure}
\centering
\includegraphics[scale=0.5]{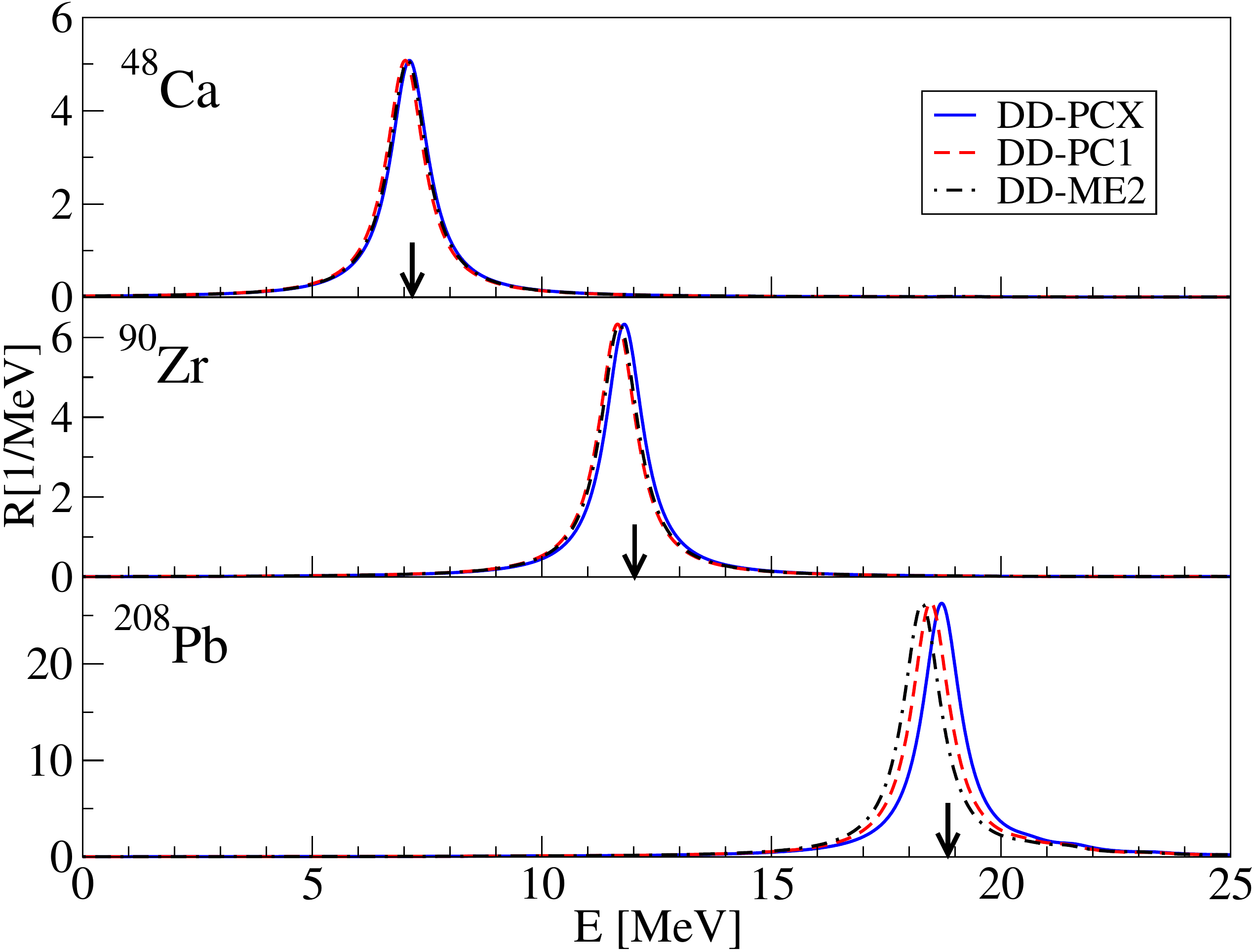}
\caption{The PN-RRPA  isobaric analog resonance transition strength distribution for
$^{48}$Ca, $^{90}$Zr and $^{208}$Pb, calculated using DD-PCX, DD-PC1, and DD-ME2 functionals.}
\label{ias_closed}
\end{figure}

The PN-R(Q)RPA based on point coupling interactions introduced in 
Sec. \ref{secII} is first implemented in the case of $J^{\pi}=0^{+}$ charge-exchange 
transition, isobaric analog resonance (IAR). It is induced by the Fermi isospin-flip operator,
\begin{equation}
\hat{T}_{\beta^{\pm}}^{F}=\sum_{i=1}^{A}\tau_{\pm}.
\label{iaroperator}
\end{equation}
Figure~\ref{ias_closed} shows the transition strength distributions for
the IAR in closed shell nuclei $^{48}$Ca, $^{90}$Zr and $^{208}$Pb, 
calculated with the PN-RRPA using two density-dependent point coupling interactions,
DD-PCX and DD-PC1, and density-dependent meson-exchange effective
interaction DD-ME2. As expected, for each nucleus the response to the Fermi
operator results in a pronounced single IAR peak. The IAR peak energy and transition strength
display rather moderate model dependence. The most pronounced spread
of the IAR excitation energies for different interactions, about 1 MeV, is obtained 
for the heaviest system, $^{208}$Pb, while for $^{48}$Ca and $^{90}$Zr differences
are smaller. The results of model calculations are compared with
the experimental data for IAR excitation energies, denoted by arrows, obtained
from $(p,n)$ scattering on $^{48}$Ca~\cite{And.85}, $^{90}$Zr~\cite{Bai.80,Wakasa1997}, 
and $^{208}$Pb~\cite{Aki.95}. Good agreement of the PN-RRPA results
with the experimental data is obtained. In all three cases the calculated transition 
strengths of the IAR fulfil the Fermi non-energy weighted sum rule, consistent 
with Ref. \cite{Paar2004}.

\begin{figure}
\centering
\includegraphics[scale=0.5]{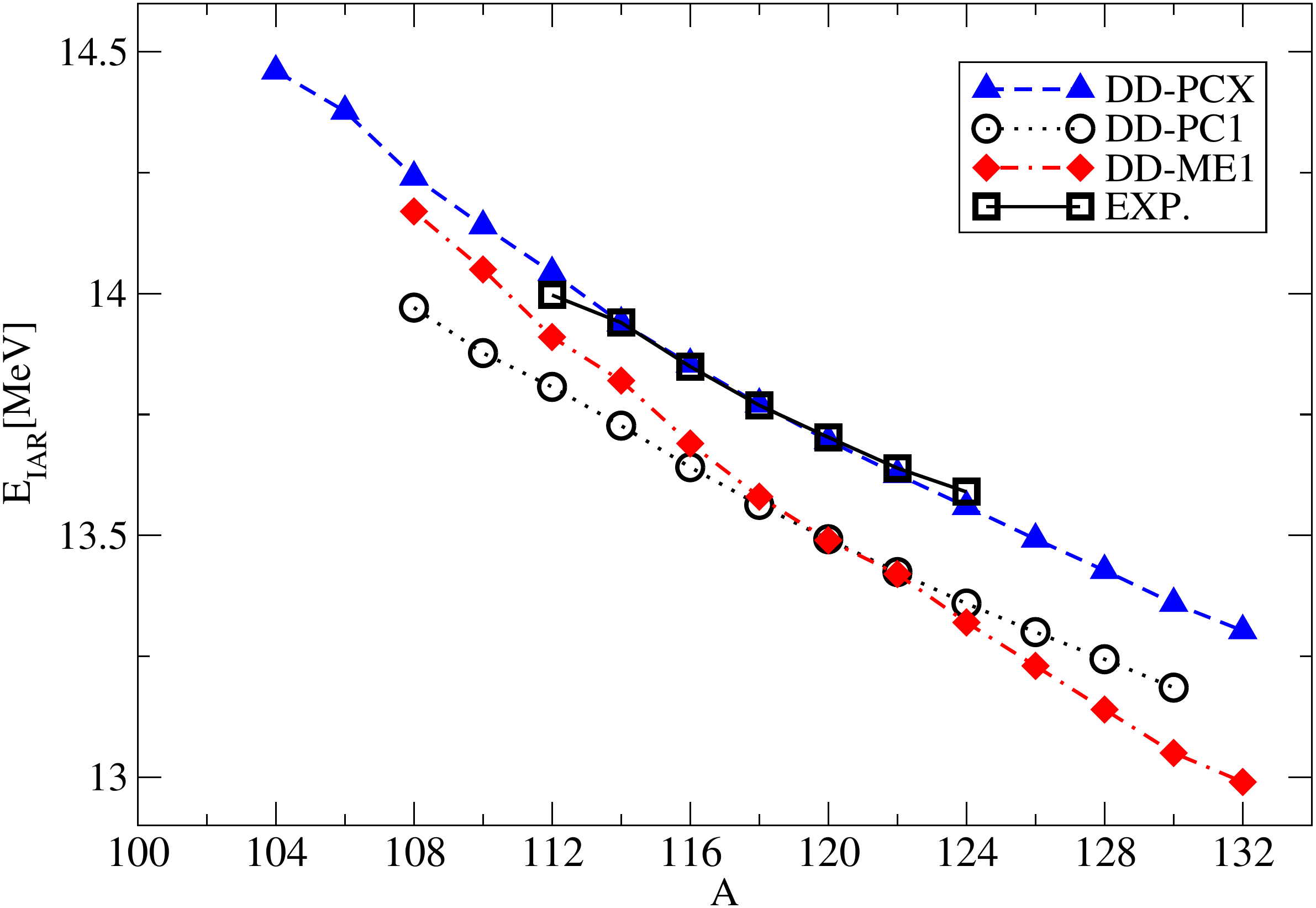}
\caption{The PN-RQRPA isobaric analog resonance excitation energy for the isotope chain $^{104-132}$Sn, with the  proton-neutron pairing included. Calculations are based on the point coupling interaction DD-PCX and DD-PC1.  The results from the previous study based on DD-ME1 interaction \cite{Paar2004}  and the experimental data from Ref. \cite{Pham1995} are shown for comparison.}
\label{iar_sn_isotopes}
\end{figure}

Next we explore the evolution of the IAR within the Sn isotope chain, for
$A=$104$-$132. In Fig. \ref{iar_sn_isotopes} the IAR transition strength
distributions are shown for representative cases, $^{112,116,122,130}$Sn.
Model calculations are based on the PN-RQRPA with DD-PCX interaction.
The results without and with the $T=1$ proton-neutron pairing in the residual 
PN-RQRPA interaction are shown separately, in comparison to the experimental 
data from a systematic study of the ($^3$He,t) charge-exchange 
reaction in stable Sn isotopes \cite{Pham1995}. As shown in Fig. \ref{iar_sn_isotopes},
the full PN-RQRPA calculations result in a pronounced single IAR peak, with the
excitation energy that is in excellent agreement with the experimental data for
$^{112,116,122}$Sn \cite{Pham1995}. Complete treatment of paring correlations
both in the RHB and PN-RQRPA is essential for description of the IAR \cite{Paar2004}.
This is illustrated in Fig. \ref{iar_sn_isotopes}, where the strength functions 
are also shown without including the $T=1$ proton-neutron residual pairing interaction,
i.e., only the  the $ph$-channel of the RQRPA residual interaction is included.
Without  the contributions of the $pp$-channel, pronounced fragmentation of the 
transition strength is obtained for $^{112,116}$Sn, and the excitation energies
are overestimated. By including the
attractive proton-neutron pairing interaction, the transition strength becomes 
redistributed toward a single pronounced IAR peak, that is consistent with the
expectation of a narrow resonance peak from the experimental study \cite{Pham1995}. 
More pronounced effect of the residual pairing interaction is obtained
for $^{112,116,122}$Sn, while for $^{130}$Sn that is near the neutron 
closed shell the effect is rather small.
\begin{figure}
\centering
\includegraphics[scale=0.5]{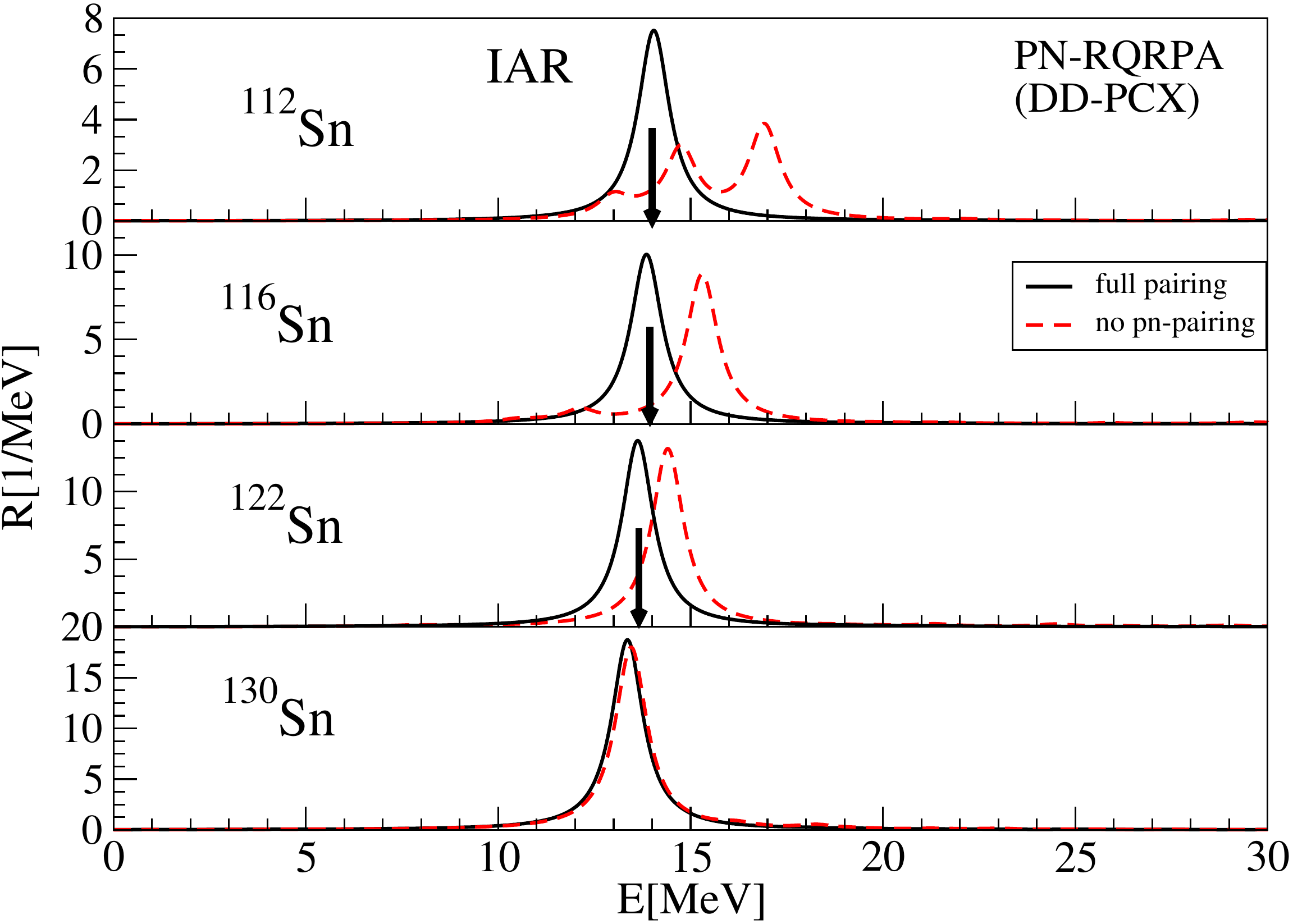}
\caption{The isobaric analog resonance transition strength distribution for $^{112,116,122,130}$Sn, calculated with the PN-RQRPA using DD-PCX interaction.The results without (dashed line) and with (solid line) proton-neutron pairing in the residual PN-RQRPA interaction are shown separately, in comparison to the experimental data from Ref. \cite{Pham1995}, denoted by arrows.}
\label{iar_sn}
\end{figure}

Figure \ref{iar_sn_isotopes} shows the evolution of the IAR excitation energy
for the isotopic chain $^{104-132}$Sn, with the  proton-neutron pairing
included in the PN-RQRPA. The results are shown for the point coupling 
interactions DD-PCX and DD-PC1.  For comparison, the IAR excitation energies
from the previous study based on DD-ME1 interaction \cite{Paar2004}  and the 
experimental data from Ref. \cite{Pham1995} are displayed.
As one can observe in the figure, recently established interaction DD-PCX
reproduces the experimental data with high accuracy, while the DD-PC1 and
DD-ME1 interactions provide systematically lower energies. We note that
DD-PCX parameterization has been established using additional constraints
on nuclear collective transitions that resulted with improved isovector
properties, essential for the description of nuclear ground state, excitation phenomena,
and nuclear matter properties around the saturation density \cite{Yuksel.09}.
Clearly, the PN-RQRPA framework based on DD-PCX interaction introduced
in this work represents a considerable progress in comparison to other 
approaches.

%========================================================================= 
\subsection{The Gamow-Teller resonance}

%=========================================================================
% 

\begin{figure}
\centering
\includegraphics[scale=0.5]{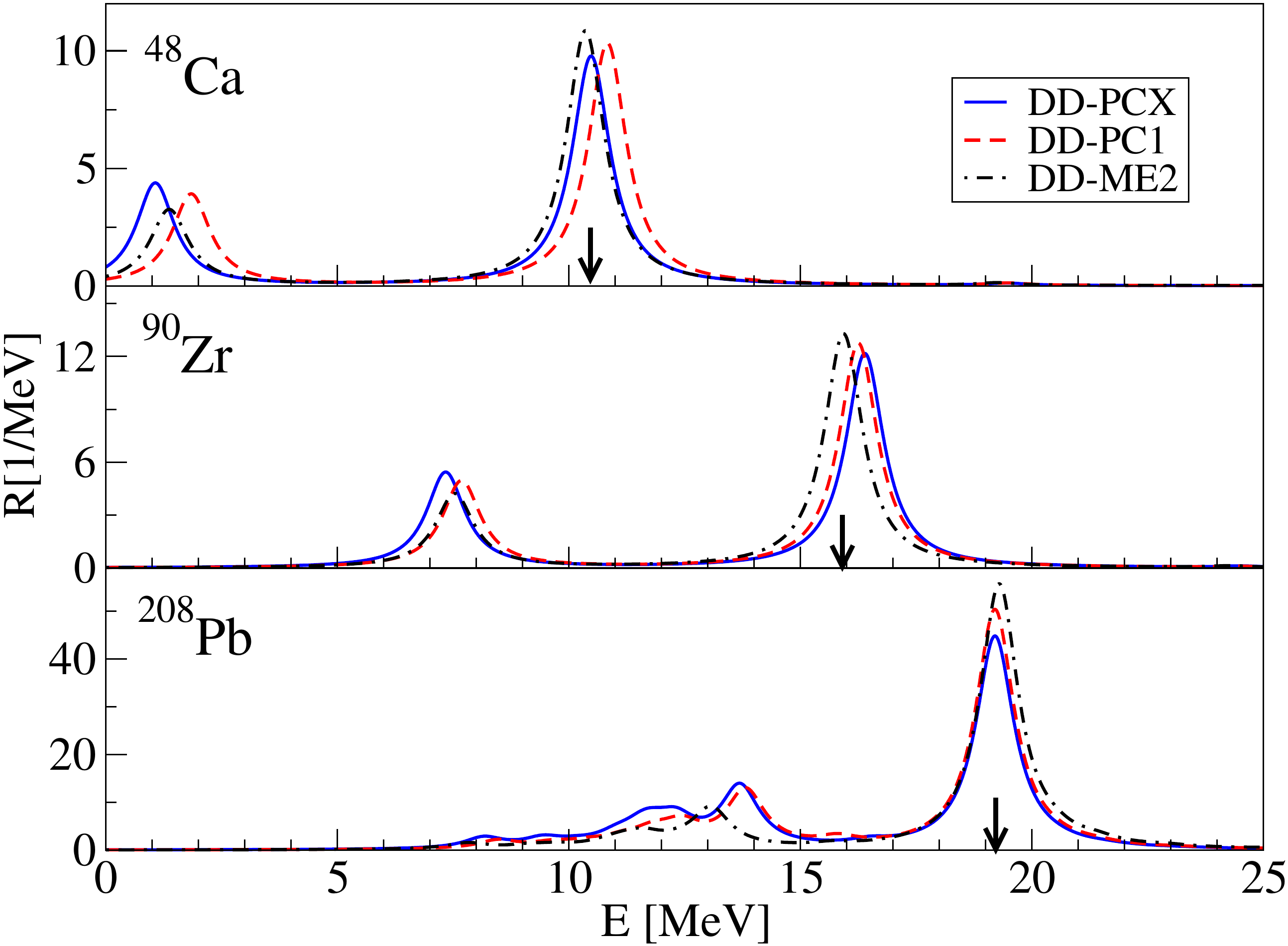}
\caption{The GT$^-$ strength distribution for $^{48}$Ca, $^{90}$Zr and $^{208}$Pb, calculated
using the DD-PCX, DD-PC1, and DD-ME2 functionals. The experimental values
of the main  GT$^-$ peak for $^{48}$Ca~\cite{And.85}, $^{90}$Zr~\cite{Wakasa1997} and $^{208}$Pb~\cite{Aki.95,Pham1995,Horen1980} are denoted with arrows.}
\label{fig_gt}
\end{figure}

The Gamow-Teller transitions involve both the spin and isospin degrees
of freedom. In the charge-exchange excitation spectra, these transitions 
mainly concentrate in a pronounced resonance peak - 
Gamow-Teller resonance (GTR), representing a coherent superposition of 
$J^{\pi}=1^{+}$ proton-particle -- neutron-hole transitions
of neutrons from orbitals with $j = l + \frac{1}{2}$ into protons in 
orbitals with $j = l - \frac{1}{2}$. The GT transitions are excited by
the spin-isospin operator
\begin{equation}
T_{\beta^{\pm}}^{GT}=\sum_{i=1}^{A}\bm{\Sigma}\tau_{\pm} \; .
\label{gtopera}
\end{equation}
Figure \ref{fig_gt} shows the GT$^-$ transition strength distribution for closed shell nuclei
 $^{48}$Ca, $^{90}$Zr and $^{208}$Pb, calculated with the PN-RQRPA using DD-PCX and DD-PC1 interactions.
For comparison, the results of the meson exchange functional with the DD-ME2 parameterization are 
also shown. The experimental values for the main GT$^-$ peak are denoted with arrows  for
$^{48}$Ca~\cite{And.85}, $^{90}$Zr~\cite{Bai.80,Wakasa1997}, and $^{208}$Pb~\cite{Horen1980,Aki.95, Kra.01}.
Since the PN-RQRPA excitation energies are given with respect to the mother nucleus, the
experimental values given with respect to daughter nucleus are modified by adding the mass difference
between the daughter and mother isotopes as well as the mass difference between neutron and proton 
which is missing when the GT is built on the basis of p-h excitations \cite{Col.94,Yuksel2020}.
The transition strength distributions are dominated by the main
Gamow-Teller resonance peak, that is composed from
direct spin-flip transitions,
($j = l + \frac{1}{2}$ $\rightarrow$ $j = l - \frac{1}{2}$). In addition, 
pronounced low-energy GT$^-$ strength is obtained,
composed from the core-polarization spin-flip
($j = l \pm \frac{1}{2}$ $\rightarrow$ $j = l \pm \frac{1}{2}$), 
and back spin-flip transitions
($j = l - \frac{1}{2}$ $\rightarrow$ $j = l + \frac{1}{2}$).
As it is well known, quantitative description of the low-energy GT$^-$ strength
is essential in modelling beta decay half-lives \cite{Marketin2007,Marketin2016}.
When using three different effective interactions as shown in Fig. \ref{fig_gt},
the spread of values of the GT$^-$ excitation energies within $\approx$ 1 MeV is 
obtained.

\begin{table}[h]
\caption{Calculated and experimental GT$^-$ properties, strength contributions for both GT$^{-}$ and GT$^{+}$ channel, i.e. $\sum B(\text{GT}^{-})$ and $\sum B(\text{GT}^{+})$, and total transition strengths $\sum B(\text{GT}^{-})-B(\text{GT}^{+})$ for $^{48}$Ca, $^{90}$Zr and $^{208}$Pb, for DD-PCX and DD-PC1 interactions. The total transition strengths are given in percentages of the Ikeda sum rule value $3(N-Z)$ \cite{Ikeda1963}, and contributions from the negative energy states of the Dirac sea are shown.
The experimental values for $^{48}$Ca} are from Ref. \cite{And.85, Yako2009}, $^{90}$Zr from Refs. \cite{Ban80, Wak97, Ray1990, CONDE1992}, and $^{208}$Pb from Ref. \cite{Aki1995,Kras2001}.
\begin{center}
\begin{tabular}[c]{ c c c c c   } 
\hline
& & $^{48}$Ca & $^{90}$Zr & $^{208}$Pb\\
\hline
Experiment & $E_1$ ($E_x$) [MeV] & 8.3 ($\approx 10.5$ \cite{And.85}) & 15.5 & 19.2 \\
 & FWHM$_1$ & 1.5 & 3.8 & 4.1 \\
 & $E_2$ [MeV] & 10.9 & - & -\\
 & FWHM$_2$ & 3.9 & - & -\\
 & $\sum B(\text{GT}^{-}+\text{IVSM}^{-})$ & $15.3\pm2.2$ \cite{Yako2009}& $34.2\pm1.6$ \cite{Wak97}& - \\
 & $\sum B(\text{GT}^{-})$ & - & $28.0\pm1.6$\cite{Wak97}  & - \\
 & $\sum B(\text{GT}^{+}+\text{IVSM}^{+})$ & $2.8\pm0.3$ \cite{Yako2009} & - & -  \\
 & $\sum B(\text{GT}^{+})$ & $1.9\pm0.5$ \cite{Yako2009} &  \begin{tabular}{c} $1.0\pm0.3$\cite{Ray1990} \\($1.7\pm0.2$)\cite{CONDE1992} \end{tabular}  & - \\ 
 & $3(N-Z)$ & 24 & 30 & 132\\
 & $\sum B(\text{GT}^{-})-B(\text{GT}^{+})$ [\%] & \begin{tabular}{c} $(52\pm9)$\% \\($\approx70\%$ \cite{And.85})\end{tabular} & $(90\pm5)\%$ \cite{Wak97} & $60-70\%$ \cite{Aki1995}\\
 \hline
DD-PCX & $E_x$ [MeV] & 10.61 & 16.73 & 19.21 \\
 & $\sum B(\text{GT}^{-})$ & 27.34 & 38.99 & 147.33\\
 & $\sum B(\text{GT}^{+})$ & 3.34 &  8.89 & 15.33 \\
 & $\sum B(\text{GT}^{-})-B(\text{GT}^{+})$ [\%] & 99.96\% & 100.33\% & 99.99\% \\
 & Dirac sea [\%] & 6.48\% & 7.90\% & 8.23\% \\
 \hline
 DD-PC1 & $E_x$ [MeV] & 10.98 & 16.54 & 19.21 \\
 & $\sum B(\text{GT}^{-})$ & 27.09 & 37.21 & 145.92 \\
 & $\sum B(\text{GT}^{+})$ & 3.10 & 7.05 & 13.93\\
 & $\sum B(\text{GT}^{-})-B(\text{GT}^{+})$ [\%] & 99.97\% & 100.50\% &  99.99\%\\
 & Dirac sea [\%] & 5.93\% &  7.46\% & 7.48\% \\
 \hline\hline
\end{tabular}
\end{center}
\label{TabGTmagic}
\end{table}

While the strength parameter of the pseudovector channel in the residual
PN-R(Q)RPA interactions is constrained by the GTR excitation energy
in $^{208}$Pb, reasonable agreement with experimental data is 
obtained for $^{48}$Ca and $^{90}$Zr without additional adjustments
of the effective interaction. As it has already been discussed in previous
studies, the Ikeda sum rule for GT transition strength \cite{Ikeda1963} is fully reproduced
in a complete calculation that includes both the configurations formed 
from occupied states in the Fermi sea and empty negative-energy 
states in the Dirac sea ~\cite{Kur.03a,Kur.03b,Ma.03,Kur.04,Paar2004}. 
Table \ref{TabGTmagic} shows
the summary of the GTR properties for $^{48}$Ca, $^{90}$Zr and $^{208}$Pb for DD-PCX and DD-PC1 interactions: 
centroid excitation energies, the total transition strength difference $\sum B(\text{GT}^{-})-B(\text{GT}^{+})$ in comparison to the 
Ikeda sum rule \cite{Ikeda1963} (in $\%$), contributions of the Dirac sea states
to the sum rule (in $\%$), and the respective experimental values \cite{And1985,Ban80, Wak97,Aki1995, Yako2009}. While the 
calculations accurately exhaust the Ikeda sum rule values,
obtained total transition strengths in $^{48}$Ca and $^{90}$Zr for GT$^{-}$ channel are somewhat larger than experimental ones for both point coupling interactions. After subtraction of the estimated IVSM contribution (usually $\sim 10\%$) the values of total GT$^{-}$ strengths are even lower \cite{Yako2009}. In the GT$^{+}$ channel for $^{48}$Ca the difference between theoretical and experimental values is reduced, i.e. $\sum B(\text{GT}^{+})=3.3$ ($3.0$) for DD-PCX (DD-PC1) and it is very close to the experimental value of $\sum B(\text{GT}^{+}+\text{IVSM}^{-})=2.8$ ($\sum B(\text{GT}^{+})=1.9$ without IVSM). However, the experimental strengths in $^{48}$Ca may be significantly underestimated for higher excitation energies above 15 MeV in the GT$^{-}$ spectrum and above 8 MeV in the GT$^{+}$ spectrum \cite{Yako2009}. The largest discrepancy in the GT$^{+}$ channel is observed for $^{90}$Zr, where calculated values are at least few times greater than the highest estimates of experimental GT$^{+}$ strengths\cite{CONDE1992}. Other nonrelativistic RPA approaches, such as Extended RPA theories account also somewhat larger value of $\sum B(\text{GT}^{-})$ (usually 10-20\%) in lower part of $^{48}$Ca excitation spectrum ($E_x \lesssim 20$ MeV)\cite{Rij1993}. Both dressed and extended RPA theories overestimate total experimental strengths in $^{90}$Zr for GT$^{-}$ channel by $20-60\%$ for $E_x\lesssim 25$ MeV. However, in these calculations significant percentage of the sum rule for both GT$^{-}$ and GT$^{+}$ may be found for excitation energies above $E_x\gtrsim$ 40 MeV \cite{Rij1993}, which is not the case in our calculations. The nonrelativistic QRPA + PVC calculations also overestimate the total experimental strengths in  $^{48}$Ca (71\% of the RPA+PVC strength) and $^{208}$Pb (63\% of the RPA+PVC strength) \cite{Niu2014PVC}, still not providing the explanation for the missing experimental strength.
The contributions of the GT transitions to the empty Dirac sea in doubly magic nuclei is from $6.5-8.2\%$ ($5.9-7.5\%$) of the total strength for DD-PCX (DD-PC1), in agreement with the previous studies \cite{Paar2004}.

\begin{figure}
\centering
\includegraphics[scale=0.5]{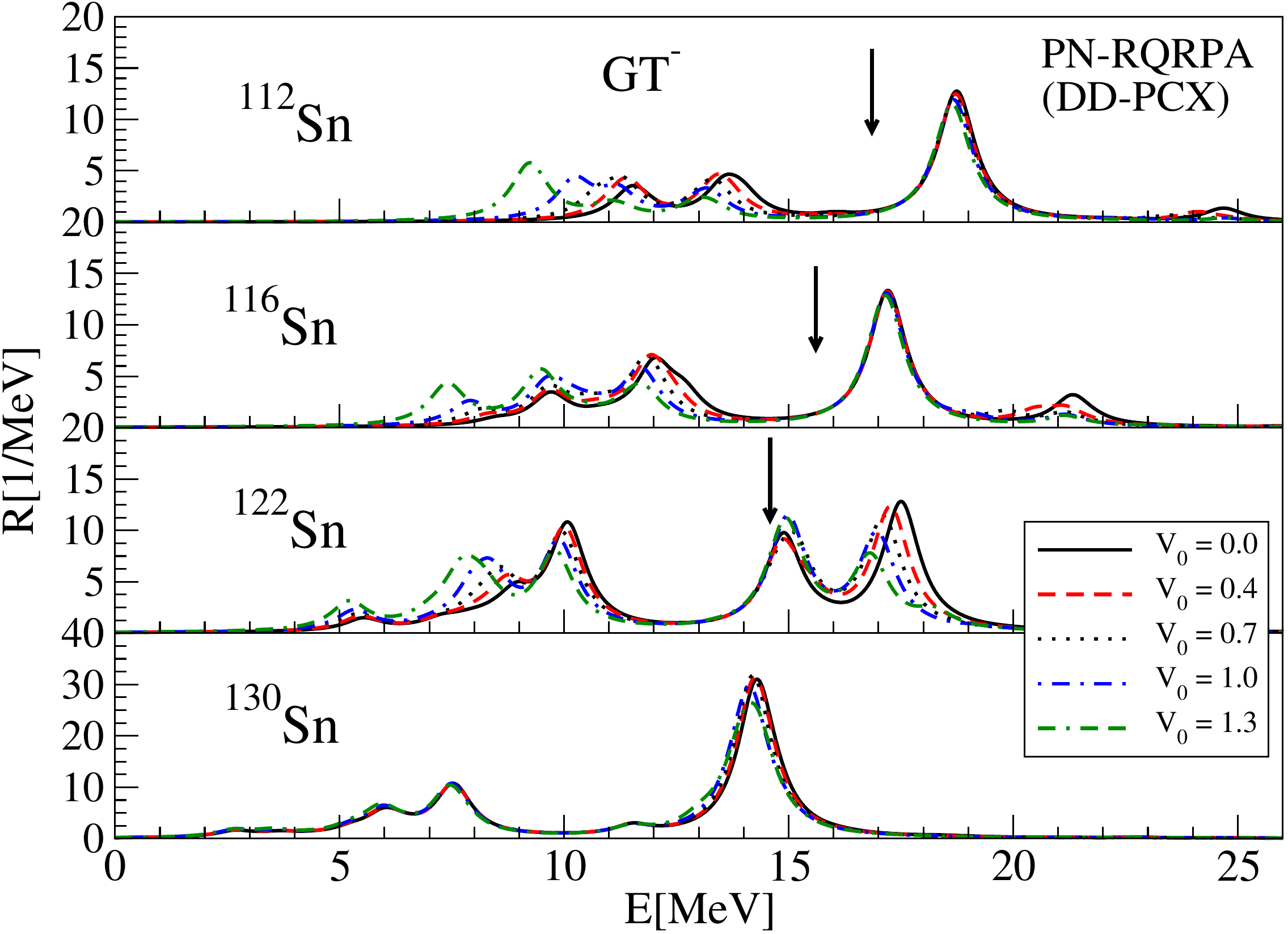}
\caption{The GT$^-$ strength distribution for $^{112,116,122,130}$Sn, calculated with the PN-RQRPA using DD-PCX interaction, for the range of values of the isoscalar proton-neutron pairing interaction strength parameter, $V_0$=0-1.3. Experimental values for the main GT$^{-}$ peaks for $^{112}$Sn, $^{116}$Sn and $^{122}$Sn are denoted with arrows \cite{Pham1995}.}
\label{gtr_sn}
\end{figure}
In the following, the PN-RQRPA based on point coupling interaction DD-PCX
is employed in the study of GT$^-$ transitions in Sn istotope chain. For open
shell nuclei, in addition to the separable pairing interaction included in the RHB,
the PN-RQRPA residual interaction also includes the isoscalar proton-neutron 
pairing as introduced in Sec. \ref{secII}. Since this pairing interaction channel
is not present within the RHB, its strength parameter $V_0$ can be constrained 
by the experimental data, e.g., on GT$^-$ excitation energies or beta decay
half-lives. Rather than providing the optimal value of $V_0$, in the present
analysis we explore the pairing properties and sensitivity of the GT$^-$ 
transitions by systematically varying $V_0$. In Fig.\ref{gtr_sn}, the 
GT$^-$ strength distribution is shown for the isotopes $^{112,116,122,130}$Sn, 
calculated with the PN-RQRPA using DD-PCX interaction. The 
isoscalar proton-neutron pairing interaction strength parameter is varied within
the range of $V_0$=0-1.3 MeV. At higher excitation 
energies, a pronounced GT resonance peak is obtained, 
except for $^{122}$Sn, where the main peak is split. In all cases, pronounced
low-energy GT$^-$ strength is obtained, spreading over the energy range
of $\approx$ 5 MeV. While the direct spin-flip transitions 
($\nu j = l + \frac{1}{2}$ $\rightarrow$ $\pi j = l - \frac{1}{2}$)
dominate the high-energy region, the low-energy strength
is dominated by core-polarization spin-flip 
($\nu j = l \pm \frac{1}{2}$ $\rightarrow$ $\pi j = l \pm \frac{1}{2}$), 
and back spin-flip 
($\nu j = l - \frac{1}{2}$ $\rightarrow$ $\pi j = l + \frac{1}{2}$)
transitions. Considering the dependence of the GT$^-$ strength on the
$T=0$ proton-neutron pairing, one can observe in Fig.\ref{gtr_sn}
that the main GTR peak appears insensitive to this pairing interaction
channel. In the low-energy region,
pronounced sensitivity of the GT$^-$ spectra on $V_0$ is obtained, i.e.
with increased $V_0$ the strength is shifted toward lower energies.
This result is in agreement with previous studies based on different
effective interactions both in the $ph$ and $pp$ channels \cite{Paar2004}.

\begin{table}[h]
\caption{Calculated values of GTR properties for $^{112}$Sn, $^{116}$Sn, $^{122}$Sn and $^{130}$Sn with DD-PCX and DD-PC1 interactions and $T=0$ pairing strengths $V_{0}~=~0.7$ and $V_{0}~=~1.3$. The contributions of strengths in the GT$^{-}$ and GT$^{+}$ channels are shown in separate columns. The total strength of GTR is also represented in percentages of the Ikeda sum rule value, $3(N-Z)$ \cite{Ikeda1963}. Contributions to total GTR strength from transitions to negative energy states of Dirac sea are also shown.}
\begin{center}
\renewcommand{\arraystretch}{1.0}
\begin{tabular}[c]{c c c c c c c c  } 
\hline
\multirow{3}{*}{} &
\multirow{3}{*}{} &
\multirow{3}{*}{$3(N-Z)$} &
  \multicolumn{3}{c}{ } &
  \multicolumn{2}{c}{Sum rule \%}\\
& & & $\sum B(\text{GT}^{-})$ & $\sum B(\text{GT}^{+})$ & $\sum \left(B(\text{GT}^{-}) - B(\text{GT}^{+})\right)$ & total & Dirac sea \\
 \hline
 DD-PCX & $^{112}$Sn & 36 & 47.22 & 11.68 &  35.53 & 98.69\% & 7.36\%\\
$(V_{0} = 0.7)$ & $^{116}$Sn & 48 & 57.77 & 9.99 & 47.77 & 99.53\% & 7.84\%  \\
 & $^{122}$Sn & 66 & 75.26 & 9.34 & 65.91 & 99.87\% & 7.87\% \\
 & $^{130}$Sn & 90 & 99.06 & 9.12 & 89.94 & 99.94\%  & 7.76\%\\
\hline
DD-PC1 & $^{112}$Sn & 36 & 45.55 & 9.84 & 35.71 & 99.21\%  & 7.11\% \\
$(V_{0} = 0.7)$ & $^{116}$Sn & 48 & 57.03 & 9.26 & 47.77 &  99.53\% & 7.85\%\\
 & $^{122}$Sn & 66 & 74.41 & 8.47 & 65.94 & 99.91\% & 7.17\% \\
 & $^{130}$Sn & 90 & 98.23 & 8.26 & 89.96 & 99.96\% & 7.01\%  \\
\hline
DD-PCX & $^{112}$Sn & 36 & 45.68 & 10.15 & 35.52 & 98.69\% & 7.36\% \\
$(V_{0} = 1.3)$ & $^{116}$Sn & 48 & 56.79 & 7.59 & 49.20 & 99.53\% & 7.85\%  \\
 & $^{122}$Sn & 66 & 74.96 & 9.04 & 65.91 & 99.87\% & 7.87\% \\
 & $^{130}$Sn & 90 & 99.01 & 9.06 & 89.94 & 99.94\%  & 7.76\%\\
 \hline
DD-PC1 & $^{112}$Sn &36 & 44.67 & 8.95 & 35.71 & 99.21\% & 7.12\% \\
$(V_{0} = 1.3)$ & $^{116}$Sn & 48 & 56.29 & 8.46 & 47.82 & 99.64\% & 7.20\% \\
 & $^{122}$Sn & 66 &74.26 & 8.32 & 65.94 & 99.91\% & 7.18\% \\
 & $^{130}$Sn & 90 & 98.20 & 8.23 & 89.96 & 99.96\% & 7.02\% \\
 \hline\hline
\end{tabular}
\label{SnSumRule}
\end{center}
\end{table}
In Tab. \ref{SnSumRule} the calculated GT strengths for the $^{112,116,122,130}$Sn isotopes
are shown for DD-PCX and DD-PC1 functionals, using $T~=~0$ strength parameter set to $V_{0}~=~0.7$ and $V_{0}~=~1.3$. The Ikeda sum rule \cite{Ikeda1963} is reasonably well reproduced.

In Tab. \ref{SnGTpeak} calculated values of moments $m_0(\text{GT}^-)$ are shown for the first three dominant GT$^{-}$ peaks,
with the corresponding excitation energies $m_1$/$m_0$ with respect to the IAS for $^{112}$Sn, $^{116}$Sn, $^{122}$Sn and $^{130}$Sn.
Calculations include the DD-PCX functional and $T=0$ pairing strength $V_{0}~=~1.3$. 
We note that there are few methods one can use to constrain the value of parameter $V_0$. One of them is based on comparison of experimental and theoretical values of relative positions of the GT peaks with respect to the IAS. As shown in Fig. \ref{gtr_sn}, the number, position and intensity of GT peaks strongly depend on the value of parameter $V_0$. In the GTR spectrum of Sn isotopes one can observe two less intensier distinguishable peaks for $V_0\lesssim0.4$ in the lower part of spectrum (GT2 and (GT3), besides one (GT1) or two dominant peaks (GT1a and GT1b), which is characteristic for $A\gtrsim122$. For values $V_0\gtrsim 0.4$ usually the third peak (GT4) starts to show up, while for values $V_0\gtrsim 0.7$ we have in general three smaller peaks in lower part of spectrum that may be distinguished. The position of these peaks and respective transition strengths are strongly influenced by $T~=~0$ pairing, i.e. with higher strength $V_0$ their positions move towards lower energies. Furthermore, the strength distribution is also modified, i.e. GT3 and GT4 peaks become more dominant than the GT2 ones for values $V_0\gtrsim1.3$. The influence of $T~=~0$ pairing becomes suppressed when approaching the magic neutron number N=82, as observed in the GTR spectrum of $^{130}$Sn. Experimental values of GT positions and cross sections obtained from  Sn($^3$He,t)Sb reactions (see Ref. \cite{Pham1995}) have some hierarchy, with descending values of cross sections corresponding to each peak as the peak position is moved toward lower energies, with minor exception of GT4 which intensity is similar to the GT3 one or somewhat larger. Therefore we may impose some constraints on the value of $V_0$, which should be at least $\approx0.7$ in order to reproduce the GTR properties observed in experiments.

\begin{table}[h]
\caption{
Calculated values of moments $m_0$ and corresponding energy $m_1/m_0$ with respect to the IAS  for the first four dominant GT$^{-}$ peaks for $^{112}$Sn, $^{116}$Sn, $^{122}$Sn and $^{130}$Sn. The DD-PCX functional and $T=0$ pairing strengths $V_{0}~=~0.7$ and  $V_{0}~=~1.3$ are used. 
The experimental results for the difference between GTR and IAS, i.e. $\Delta^{\text{exp}}~=~E_{\text{GTR}}^{\text{exp}}-E_{\text{IAS}}^{\text{exp}}$ and differential cross sections ${d\sigma}/{d\Omega}$ are taken from Ref. \cite{Pham1995}.
${}^{\dagger}$For $^{122}$Sn, the moments $m_0$ and $m_1/m_0$ contain contributions from two most dominant peaks.
}
\begin{center}
\begin{tabular}[c]{  c c c c c c c c c c} 
 \hline
 \multirow{2}{*}{} & 
 \multicolumn{3}{c}{ } &
 \multicolumn{3}{c}{--------- $V_0~=~0.7$ ---------} &
 \multicolumn{3}{c}{--------- $V_0~=~1.3$ ---------}\\
 &  & $\Delta^{\text{exp}}$ & ${d\sigma}/{d\Omega}$ & $m_0$ & $m_1/m_0$ &  $\left(\frac{m_1}{m_0}-E_{\text{IAS}}^{\text{teo}}\right)$ & $m_0$ & $m_1/m_0$ &  $\left(\frac{m_1}{m_0}-E_{\text{IAS}}^{\text{teo}}\right)$ \\
 & & [MeV] & [mb/sr] &  & [MeV] &  [MeV] &   & [MeV] &  [MeV] \\ \hline
GT1 & $^{112}$Sn  & 2.78 & 12.4 & 19.11 & 18.70 & 4.85 & 17.69 & 18.63 & 4.78 \\
 & $^{116}$Sn & 1.68 & 16.8 & 20.51 & 17.19 & 3.45 & 19.71 & 17.15 & 3.41\\
 & $^{122}$Sn$^\dagger$  & 1.01 &  21.9 & 31.68 & 16.71 & 3.14 & 29.62 & 15.82 & 2.25\\
 & $^{130}$Sn & - & - & 50.03 & 14.22 & 0.83 & 48.55 & 14.12 & 0.73\\
 \hline
GT2 & $^{112}$Sn & -2.08 & 4.9 & 6.30 & 13.27 & -0.58 & 3.69 & 13.05 & -0.80  \\
 & $^{116}$Sn &  -3.32 & 6.2 & 10.02 & 11.85 & -1.89 &  6.54 & 11.61 & -2.13\\
 & $^{122}$Sn &  -4.59 & 9.2 & 14.26 & 9.94 & -3.63 & 11.27 & 9.78 & -3.79 \\
 & $^{130}$Sn & - &  - & 15.75 & 7.53 & -5.86 & 16.07 & 7.40 & -5.99\\
  \hline
GT3 & $^{112}$Sn &  -3.67 & 1.5 & 5.17 & 11.31 & -2.54 & 2.54 & 11.06 & -2.79\\
 & $^{116}$Sn &   -5.18 & 2.1 & 3.04 & 10.64 & -3.10 & 8.71 & 9.52 & -4.22 \\
 & $^{122}$Sn &   -7.87 & 6.3 & 10.27 & 8.33 & -5.24 & 12.81 & 7.87 & -5.70\\
 & $^{130}$Sn &  - & - & 9.74 & 5.87 & -7.52 & 10.90 & 5.80 & -7.59\\
  \hline
GT4 & $^{112}$Sn & -4.83 & 2.0 & 3.14 & 10.71 & -3.14 & 8.94 & 9.01 & -4.84\\
 & $^{116}$Sn &  -6.52 & 2.5 & 2.05 & 8.12 & -5.62 & 6.65 & 7.33 & -6.41\\
 & $^{122}$Sn &  -9.79 & 1.2 & 3.32 & 5.86 & -7.71 & 5.64 & 5.59 & -7.98\\
 \hline\hline 
\end{tabular}
\end{center}
\label{SnGTpeak}
\end{table}

\begin{figure}
\centering
\includegraphics[scale=0.5]{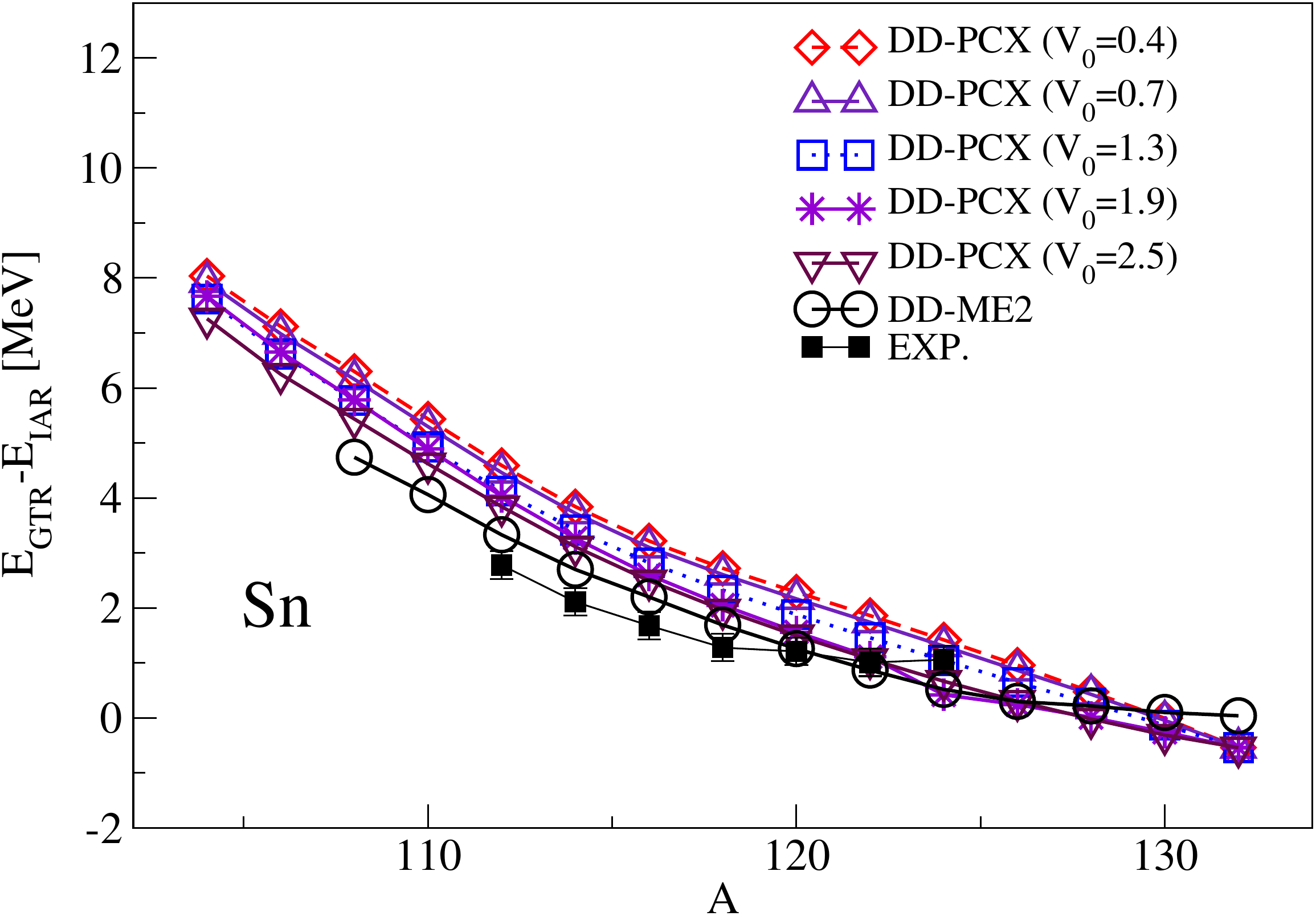}
\caption{The PN-RQRPA excitation energy for GT$^-$ direct spin-flip transitions for  $^{104-132}$Sn, for the range of values of the isoscalar proton-neutron pairing interaction strength parameter $V_0$. Calculations are based on the density dependent point coupling interaction DD-PCX and the results from the previous study based on the DD-ME2 interaction \cite{Paar2004} and the experimental data from Ref. \cite{Pham1995} are shown for comparison.}
\label{gt-ias_sn_isotopes}
\end{figure}

In Fig. \ref{gt-ias_sn_isotopes} the PN-RQRPA results for the GT$^-$ direct 
spin-flip transition excitation energy centroid with the respect to the IAR energy
are shown for the chain of even-even isotopes $^{104-132}$Sn. The available 
experimental data obtained from Sn($^3$He,t)Sb 
charge-exchange reactions are shown for comparison~\cite{Pham1995}.
The PN-RQRPA calculations are performed for the range of values of the isoscalar 
proton-neutron pairing interaction strength parameter,
$V_0$=0.4-2.5. The point coupling interaction DD-PCX is used and the 
results based on the DD-ME2 interaction are also displayed \cite{Paar2004}. We noticed almost linear decrease of differences between the GTR and IAS energies in Sn isotope chain as function of A or N (while keeping $Z=50$ fixed). Similar observations may be found in nonrelativistic calculations with Skyrme parametrizations in \cite{Fra2007}.
One can observe that the proton-neutron pairing interaction
systematically reduces the GT-IAR energy splittings along the Sn isotope chain,
resulting in good agreement for $V_0\approx$2.5. We note that relatively 
higher values of $V_0$ are required to reproduce the GT-IAR energy splittings.
In Refs.~\cite{VPNR.03, Paar2004}, it has been emphasized that the energy 
difference between the GTR and the IAS reflects the magnitude of the 
effective spin-orbit potential. As one can see in Fig. \ref{gt-ias_sn_isotopes},
the GT-IAR energy splittings reduce in neutron-rich Sn isotopes toward zero 
value, reflecting considerable reduction of the spin-orbit potential and the 
corresponding increase of the neutron skin thickness $r_n - r_p$ \cite{VPNR.03}.
Therefore, as pointed out in Ref. \cite{VPNR.03}, the energy difference
$E_{\rm GT} - E_{\rm IAS}$, obtained from the experiment, could be used to 
determine the value of neutron skin thickness in a consistent framework that
can simultaneously describe the charge-exchange excitation properties and
the $r_n - r_p$ value, such as the RHB+PN-RQRPA approach. 

In Ref. \cite{Yuk2020} the
GT$^{-}$ transition strength for $^{118}$Sn was analyzed in more details, including the finite temperature effects, using the Skyrme functional SkM$^*$.
In the present study, for $^{118}$Sn without T=0 pairing and DD-PCX interaction ($V_0$=0), we obtain pronounced low-lying GT$^{-}$ states at energies E=7.32, 8.88, 9.64, 10.21, and
11.45 MeV with the corresponding transition strengths B(GT$^{-}$)=1.21, 3.25, 1.12, 1.14,
and 14.30. The main GTR peaks are obtained at E=16.46 and 19.83 MeV with B(GT$^{-}$)=19.91 and 7.69, respectively.
The excitation energies are comparable with the respective 
values obtained for the SkM$^*$ interaction in Ref. \cite{Yuk2020}, E(GT$^{-}$)= 6.2, 8.9, 10.5, 16.3 and 20.2 MeV \cite{Yuk2020}.
Study in Ref. \cite{Yuk2020} also showed that relativistic calculations of GT$^{-}$ transitions in $^{118}$Sn with density dependent meson-exchange interaction DD-ME2 lead to the splitting between major (GT1a) and energetically higher somewhat less intensive peak (GT1b), that exists only for $V_0\lesssim 200$ MeV in $T = 0$ pairing channel. For  $V_0 \approx 200$ MeV the two peaks start to merge, while for $V_0 \gtrsim 240$ MeV they are completely replaced by one somewhat stronger peak shifted $\approx 0.5$ MeV towards higher energies.
 However, nonrelativistic calculations in Refs. \cite{Yuk2020, Fra2007} show different behavior in GT$^{-}$ spectrum of $^{118}$Sn with respect to $T = 0$ pairing strength, i.e. the two dominant peaks never merge, even for large values of overall strength of the $T = 0$ pairing ($\approx 800$ \text{MeV}). 
The same feature is obtained in the present study, as shown in Fig. \ref{gtr_sn} for $^{122}$Sn.
We note that relativistic calculation in Ref. \cite{Yuk2020} used another pairing interaction in $T = 0$ channel than the non-relativistic one, i.e. two Gaussians with different relative strengths to cover short and medium distances in coordinate space introduced separately for the PN-RQRPA. This caused additional interplay between attractive and repulsive terms which act differently for the short and medium ranges. Therefore, behavior of the GTR spectrum obtained in our study is more similar to the nonrelativistic calculations in Refs. \cite{Yuk2020, Fra2007}, where the zero range density dependent surface pairing in both isospin channels is introduced separately in the PN-RQRPA.
%

%=========================================================================

%=========================================================================
%  Section 

\section{\label{secIV}Conclusion}

In this work we have formulated a consistent framework for description of nuclear charge-exchange
transitions based on the proton-neutron relativistic quasiparticle random phase approximation in 
the canonical single-nucleon basis of the relativistic Hartree-Bogoliubov model, using
density-dependent relativistic point coupling interactions. The implementation of recently
established DD-PCX interaction\cite{Yuksel.09}, adjusted not only to the nuclear ground
state properties, but also with the symmetry energy 
of the nuclear equation of state and the incompressibility of nuclear matter constrained 
using collective excitation data, allows improved description of ground-state and 
excitation properties of nuclei far from stability, that is important for the studies of exotic nuclear
structure and dynamics, as well as for applications in nuclear astrophysics. The introduced 
formalism based on point coupling interactions is also relevant from the practical point of view, 
because it allows efficient systematic large-scale calculations of nuclear properties and processes of 
relevance for the nucleosynthesis and stellar evolution modeling. In the current formulation 
of the RHB+PN-RQRPA, the pairing correlations are implemented by the separable 
pairing force~\cite{Tian2009} that allows accurate and efficient calculations of the pairing properties.  
The PN-RQRPA includes both the $T=1$ and $T=0$ pairing channels. While the $T=1$ channel
corresponds to the pairing interaction constrained at the ground state level, $T=0$ proton-neutron
pairing strength parameter can be determined from the experimental data on charge-exchange
transitions and beta decay half lives. 
In order to validate the PN-R(Q)RPA framework introduced in this work, spin and isospin excitations
- isobaric analog resonances and  Gamow-Teller transitions - have been investigated in several 
closed-shell nuclei and Sn isotope chain. The results show very good agreement with the experimental 
data, representing an improvement compared to previous studies based on the
relativistic nuclear energy density functionals. When compared to other
theoretical approaches that usually underestimate the IAR excitation energies in Sn isotope chain, 
the present model using DD-PCX interaction accurately reproduces the experimental data.
Therefore, the framework introduced in this
work represents an important contribution for the future studies of astrophysically relevant
nuclear excitations and weak interaction processes, in particular $\beta$-decay in neutron-rich 
nuclei, electron capture in presupernova stage of massive stars, and neutrino-nucleus 
reactions relevant for the synthesis of elements in the universe and stellar evolution. 
Especially important is the extension of theory framework introduced in this work
to include both the finite temperature effects together with nuclear pairing and apply it
for the description of electron capture and beta decay at finite temperature characteristic
for stellar environment \cite{Yuksel2017,Yuksel2020}. 
This work is currently in progress \cite{Ravlic_ec2020,Ravlic2020}.

%=========================================================================

\bigskip \bigskip
%%%%%%%%%%%%%%%%%%%%%%%%%%%%%%%%%%%%%%%%%%%%%%%%%%%%%%%%%%%%%%%%%%%%%

% ACKNOWLEDGMENTS
%
\section*{Acknowledgments}
Stimulating discussions with Ante Ravli\' c are gratefully acknowledged.
This work is supported by the QuantiXLie Centre of Excellence, a project co financed by the Croatian Government and European Union through the European Regional Development Fund, the Competitiveness and Cohesion Operational Programme (KK.01.1.1.01).
Y. F. N. acknowledges support from the National Natural Science Foundation 
of China under Grant No. 12075104

\appendix
\section{\label{phelements}Particle-hole matrix elements}
For the isovector-pseudovector part of particle-hole channel in the PN-RQRPA we use the contact type of interaction 
with constant coupling,
\begin{equation}
V_{PV}=-\alpha_{PV}[\gamma_5\ \gamma^{\nu} \vec{\tau}]_1\ [\gamma_5\gamma_{\nu} \vec{\tau}]_2\  \delta(\vec{r}_1-\vec{r}_2).
\end{equation}
The uncoupled matrix element of isovector-pseudovector interaction is given by 
\begin{multline}
\langle a\ b|V|c\ d\rangle~=~-\alpha_{PV}\int dr r^2 
\sum_{L\lambda}\bigg[\int dr_1 r_1^2\int d\Omega_1 \bar{\psi}_a({r}_1,\Omega_1)[\gamma_5\ \gamma^{\nu} \vec{\tau}]_{(1)} Y_{L\lambda}^{*}(\Omega_1)
\\ 
~\delta (r -r_1)\psi_c({r}_1, \Omega_1) \bigg]\bigg[\int dr_2 r_2^2 \int d\Omega_2 \bar{\psi}_b(r_2, \Omega_2)[\gamma_5\ \gamma_{\nu} \vec{\tau}]_{(2)} Y_{L\lambda}(\Omega_2)\ \delta (r -r_2)\ \psi_d({r}_2, \Omega_2) \bigg],
\label{begPVeq}
\end{multline}
where we used expansion of delta function in spherical coordinates \cite{RingShuck},
\begin{align}
\delta(\vec{r}_1 - \vec{r}_2)~&=~\frac{\delta(r_1 - r_2)}{r_1 r_2} \sum_{\lambda \mu} Y_{\lambda \mu}^{*}(\Omega_1) Y_{\lambda \mu} (\Omega_2)\\
&=~\int d r r^2 \frac{\delta (r -r_1) \delta(r-r_2)}{r^2 r_1 r_2} \sum_{\lambda \mu} Y_{\lambda \mu}^{*}(\Omega_1) Y_{\lambda \mu} (\Omega_2).
\end{align}
Indices $a$, $b$, $c$ and $d$ refer to all quantum numbers involved, while the Dirac's conjugate is defined in standard way as
\begin{equation}
\bar{\psi}_a=\psi_a^+\ \gamma_0.
\end{equation}
If we ignore for a moment the isospin part of the wave function (corresponding quantum numbers isospin $t$ and its third projection $t_0$) one can rewrite Eq. \eqref{begPVeq} in the following form:
\begin{equation}
\langle a m_a\ b m_b | V | c m_c\ d m_d \rangle^{\text{PV}}~=-\alpha_{PV}\int dr r^2 \sum_{\lambda \mu}(-1)^\mu \sum_\nu\left[Q_{\lambda \mu}^\nu(r, \Omega_1)\right]_{ac}\left[Q_{\lambda -\mu; \nu}(r, \Omega_2)\right]_{bd}
\end{equation}
where the index $\nu$ refers to Dirac matrix $\gamma_\nu$ ($\gamma^\nu$) in the vertex and we used (see Ref. \cite{Gre.00} for the sherical case of spinor):
\begin{align}
\left[Q_{\lambda \mu}^\nu(r, \Omega_i)\right]_{am_a\ cm_c}~=&~ 
\int d \Omega_i \left( f_{n_a \kappa_a} (r) \Omega_{\kappa_a m_a}^{\dagger}(\Omega_i)~~-ig_{n_a \kappa_a}(r)\Omega_{\bar{\kappa}_a m_a}^{\dagger} (\Omega_i)\right)
[\gamma_0 \gamma_5\ \gamma^{\nu}]_{(i)}
    \nonumber\\
    &~\times Y_{\lambda \mu}(\Omega_i)
    \left(\begin{array}{clcr}
f_{n_c \kappa_c} (r) \Omega_{\kappa_a m_c}(\Omega_i)\\
ig_{n_c \kappa_c}(r) \Omega_{\bar{\kappa}_c m_c} (\Omega_i)\end{array}\right),
\label{simpleQ}
\end{align}
where the subscripts $1$ and $2$ refer to $\Omega_{\text{i = 1,2}} = \left(\theta_i, \phi_i\right)$, while radial parts of bi-spinors $f(r)$ and $g(r)$ are assumed to be real in our case. From Eq. (\ref{simpleQ}) one can easily obtain timelike
\begin{align}
        \left[Q_{\lambda \mu}^0(r, \Omega_i)\right]_{a m_a\ c m_c}~=&~-\bigg[i\ f_{n_a\kappa_a}(r)\ g_{n_c\kappa_c}(r){ \langle (1/2\ l_a)j_a m_a|\ Y_{\lambda \mu}(\Omega_i)|(1/2\bar{l_c})j_c m_c\rangle} \nonumber\\
   & -i\ g_{n_a\kappa_a}(r)\ f_{n_c\kappa_c}(r) {\langle (1/2 \bar{l_a})j_a m_a|\ Y_{\lambda \mu}(\Omega_i)|(1/2 l_c)j_c m_c\rangle} \bigg]
\label{PVq0}
\end{align}
and spacelike components ($k = 1,2$ and $3$)
\begin{align}
\left[Q_{\lambda \mu}^k(r, \Omega_i)\right]_{a m_a\ c m_c}~=&~-\bigg[ f_{n_a \kappa_a}(r)  f_{n_c \kappa_c}(r) \langle (1/2 l_a) j_a m_a | \sigma_k Y_{\lambda \mu}(\Omega_1) | (1/2 l_c) j_c m_c \rangle \nonumber\\ 
&~+~g_{n_a \kappa_a}(r)  g_{n_c \kappa_c}(r) \langle (1/2 \bar{l}_a) j_a m_a | \sigma_k Y_{\lambda \mu}(\Omega_1) | (1/2 \bar{l}_c) j_c m_c \rangle  \bigg].
\label{PVqi}
\end{align}
In order to couple particle and hole operator, i.e. states, into good total angular momentum J (and projection M) in the matrix elements of the residual particle-hole two body interaction, one needs to start from \cite{Suho.07}
\begin{align}
| (p h^{-1}) J M \rangle ~&=~ 
 \left[ c_p^\dagger h_h^\dagger \right]_{J M} | 0 \rangle\\
 &=~ \sum_{m_p m_h} C_{j_p m_p j_h m_h}^{J M} c_{p m_p}^\dagger h_{h m_h}^\dagger | 0 \rangle\\ 
  &= \sum_{m_p m_h}(-1)^{j_h - m_h} C_{j_p m_p j_h -m_h}^{J M}  a_{p m_p}^\dagger a_{h\ m_h} | 0 \rangle,
\end{align} 
from which $JJ$-coupled particle-hole matrix element follows directly
\begin{align}
\langle a c^{-1} | V | b^{-1} d\rangle^{J M}~=&~\sum_{m_a m_c} (-1)^{j_c - m_c}C_{j_a\ m_a\ jc\ -m_c}^{J\ M} \sum_{m_b m_d} (-1)^{j_b-m_b}C_{j_d\ m_d\ j_b\ -m_b}^{J\ M} \nonumber  \\
&~~\langle a m_a\ b m_b | V | c m_c\ d m_d \rangle .
\end{align}
Using the Wigner-Eckart theorem for angular parts in Eqs. \eqref{PVq0} and \eqref{PVqi} after some manipulations with Clebsch-Gordan coefficients one obtains the $ph$ matrix elements in $JJ$-coupled form. Therefore, after including isospin part of the matrix elements, which gives a factor of 2, which we have ignored for the moment, for the timelike part of pseudovector coupling we obtain:
\begin{align}
V_{abcd}^{(PVt) J}&=\frac{2 \alpha_{PV}}{2J+1}\langle (1/2 l_a) j_a|| Y_J (\Omega_1) || (1/2 \bar{l}_c)  j_c \rangle  \langle (1/2 {l}_d) j_d || Y_J (\Omega_2) || (1/2 \bar{l}_b) j_b\rangle\nonumber\\
&~~\int dr r^2 \bigg[ f_a(r)g_c(r) - g_a(r)f_c(r) \bigg] \bigg[ f_b(r)g_d(r) - g_b(r)f_d(r) \bigg]
\end{align}
and for spacelike part
\begin{align}
&V_{abcd}^{PV(s)J}~=~\frac{2 \alpha_{PV}}{2 J + 1} 
%\left(\frac{f_\pi}{m_\pi}\right)^2 
\sum_L\int dr r^2 \bigg[ f_a(r)f_c(r)\langle (1/2\ l_a) j_a ||\left[\sigma_S Y_L\right]_J|| (1/2\ l_c)j_c \rangle \nonumber\\ &~+~ g_a(r)g_c(r) \langle (1/2\ \bar{l}_a) j_a ||\left[\sigma_S Y_L\right]_J|| (1/2\ \bar{l}_c) j_c \rangle\bigg] \bigg[ f_b(r)f_d(r)\langle (1/2\ l_d) j_d || \left[\sigma_S Y_L\right]_J || (1/2\ l_b) j_b \rangle\nonumber\\ &~ +~ g_b(r)g_d(r)\langle (1/2\ \bar{l}_d) j_d || \left[\sigma_S Y_L\right]_J || (1/2\ \bar{l}_b) j_b \rangle\bigg],
\label{PVSend}
\end{align}
where the reducible angular part of matrix elements can be written in terms of 3jm symbols:
\begin{equation}
\langle (1/2l_a) j_a || Y_J || (1/2l_c) j_c \rangle ~=~ \frac{1 + (-1)^{l_a+l_c+J}}{2} \frac{\hat{j}_a\hat{j}_c\hat{J}}{\sqrt{4\pi}}(-1)^{j_a - 1/2}
\left(\begin{array}{clcr}
j_a & J & j_c\\
-1/2 & 0 & 1/2  \end{array}\right),
\end{equation}
and
\begin{multline}
\langle (1/2l_a) j_a || \left[\sigma_S Y_L \right]_J || (1/2l_c) j_c \rangle ~=~ \frac{1 + (-1)^{l_a+l_c+J}}{2} \frac{\hat{j}_a\hat{j}_c\hat{J} \hat{L}}{\sqrt{4\pi}}(-1)^{l_a + L}\Bigg[(-1)^{l_c+j_c+1/2}  \\
~
\left(\begin{array}{clcr}
1 & L & J\\
0 & 0 & 0  \end{array}\right)\left(\begin{array}{clcr}
j_a & J & j_c\\
-1/2 & 0 & 1/2  \end{array}\right) - \sqrt{2}\left(\begin{array}{clcr}
1 & L & J\\
-1 & 0 & 1  \end{array}\right)
\left(\begin{array}{clcr}
j_a & J & j_c\\
1/2 & -1 & 1/2  \end{array}\right)\Bigg].
\end{multline} 
Note that the timelike pseudovector matrix elements are non-zero only in the case of unnatural parity transitions, like Gamow-Teller transition. The isovector-vector part of two-body interaction is given by
\begin{equation}
V_{V}=\alpha_{V}\left[\rho_V(r_1)\right][\gamma^{\mu} \vec{\tau}]_1\ [\gamma_{\mu} \vec{\tau}]_2\  \delta(\vec{r}_1-\vec{r}_2).\\
\label{Vv}
\end{equation}
There are no additional rearrangement terms due to isospin restriction of the PN-R(Q)RPA, i.e. changing the nucleon from neutron to proton or \textit{vice versa} and properties of the point coupling functional itself. Therefore, one should start from Eq. \eqref{Vv} and follow the same procedure described before, from Eqs. \eqref{begPVeq} to \eqref{PVSend}, replacing Eqs. \eqref{PVq0} and \eqref{PVqi} with
\begin{align}
\left[Q^{k}_{\lambda \mu}(r, \Omega_i)\right]_{a m_a\ c m_c}~=~ \bigg[ i f_{n_a \kappa_a}(r) g_{n_c \kappa_c}(r)\langle (1/2\ l_a) j_a m_a | \sigma_k Y_{\lambda \mu}(\Omega_i) | (1/2\ \bar{l}_c) j_c m_c \rangle \nonumber\\- i g_{n_a \kappa_a}(r) f_{n_c \kappa_c}(r) \langle (1/2\ \bar{l}_a) j_a m_a | \sigma_k Y_{\lambda \mu}(\Omega_i) | (1/2 l_c) j_c m_c \rangle \bigg]
\end{align}
and
\begin{align}
\left[Q^{0}_{\lambda \mu}(r, \Omega_i)\right]_{a m_a\ c m_c}~=&~ \bigg[ f_{n_a \kappa_a}(r) f_{n_c \kappa_c}(r) \langle (1/2\ l_a) j_a m_a | Y_{\lambda \mu}(\Omega_i) | (1/2\ {l}_c) j_c m_c \rangle\nonumber\\ &~ + ~ g_{n_a \kappa_a}(r) g_{n_c \kappa_c}(r)  \langle (1/2\ \bar{l}_a) j_a m_a | Y_{\lambda \mu}(\Omega_i) | (1/2\ \bar{l}_c) j_c m_c \rangle\bigg]
\end{align}
in order to obtain $JJ$-coupled $ph$ matrix elements for this channel, i.e. spacelike part
\begin{align}
&V_{abcd}^{(TVs)J}~=~\frac{2}{2J+1}\sum_L \int dr r^2 \alpha_{TV}\left[\rho_V(r)\right] \bigg[ f_a(r)g_c(r)\langle (1/2\ l_a) j_a || \left[\sigma_S Y_L\right]_J || (1/2\ \bar{l}_c) j_c \rangle 
\nonumber
\\ &~- g_a(r)f_c(r)\langle (1/2\ \bar{l}_a) j_a || \left[\sigma_S Y_L\right]_J || (1/2\ l_c) j_c \rangle \bigg] 
\bigg[ f_b(r)g_d(r)\langle (1/2\ \bar{l}_d) || \left[\sigma_S Y_L\right]_J || (1/2\ l_b) j_b \rangle\nonumber \\
&~- g_b(r)f_d(r)\langle (1/2\ l_d)j_d || \left[\sigma_S Y_L\right]_J || (1/2\ \bar{l}_b) j_b \rangle\bigg]
\label{spacelike2}
\end{align}
and timelike part
\begin{align}
V_{abcd}^{(TVt)J}~=&~\frac{2}{2J+1} \int dr r^2 \alpha_{TV}\left[\rho_V(r)\right]\bigg[ f_a(r)f_c(r) + g_a(r)g_c(r))\bigg] \bigg[ f_b(r) f_d(r) + g_b(r) g_d(r) \bigg]\nonumber\\
 &~\times\langle (1/2\ l_a)j_a || Y_J || (1/2\ l_c)j_c \rangle \langle (1/2\ l_d)j_d || Y_J || (1/2\ l_b)j_b \rangle. 
\label{timelike2}
\end{align}
%%%%%%%%%%%%%%%%%%%%%%%%%%%%%%%%%%%%
\section{\label{secpair}Natural extension of separable pairing}
In order to extend the standard $S ~=~ 0$ ($T ~=~ 1$) pairing, which was used in the RHB and RQRPA, one needs to start from general expression for particle-particle matrix element in the uncoupled form:
\begin{equation}
V_{abcd}~=~\langle n_a (1/2 l_a)j_a~n_b (1/2 l_b)j_b|V\left(1-\hat{P}^r \hat{P}^\sigma \hat{P}^\tau \right) | n_c (1/2 l_c)j_c~n_d (1/2 l_d)j_d \rangle
\end{equation}
where $\hat{P}^r$, $\hat{P}^\sigma$ and $\hat{P}^\tau$ represent the exchange of relative spatial coordinate, spin and isospin, respectively. For action of the spin exchange operator $\hat{P}^\sigma$ we simply have
\begin{equation}
P^\sigma \left.|(1/2_{(1)} 1/2_{(2)}) S M_S \right\rangle ~=~ (-1)^{S+1} \left.|(1/2_{(1)} 1/2_{(2)}) S M_S \right\rangle, 
\end{equation}
which affects just the order of coupling, i.e. the phase $(-1)^{S+1}$ of symmetry transformation of the Clebsch-Gordan coefficient. Furthermore, one may construct $P^\sigma$ mathematically in the following way:
\begin{equation}
\hat{P}_\sigma~=~\frac{1+\vec{\sigma}_{1}\cdot\vec{\sigma}_{2}}{2},
\end{equation}
and isospin exchange operator in analogous way. However, the action of $\hat{P}^r$ affects only relative spatial coordinates,
\begin{align}
\hat{P}^r |\left. N L M_L \right\rangle~=~|\left. N L M_L \right\rangle,\\
\hat{P}^r |\left. n l m_l \right\rangle~=~(-1)^l |\left. n l m_l \right\rangle,
\end{align}
Therefore, as first step one needs to do the $LS$ recoupling in \textit{bra} and \textit{ket} independently
\begin{multline}
V_{abcd}^{\text{(pair)}J}~=~\hat{j}_a \hat{j}_b \hat{j}_c \hat{j}_d \sum_{\lambda S} \sum_{\lambda ' S'} \hat{\lambda} \hat{S} \hat{\lambda'} \hat{S'} 
\left\lbrace\begin{array}{clcr}
s_a & l_a & j_a\\
s_b & l_b & j_b \\
S &\lambda & J \end{array}\right\rbrace
\left\lbrace\begin{array}{clcr}
s_c & l_c & j_c\\
s_d & l_d & j_d \\
S' & \lambda ' & J \end{array}\right\rbrace \sum_{T M_T} \sum_{T' M_{T'}} C_{1/2 -1/2 1/2 1/2}^{T M_T}\\ C_{1/2 -1/2 1/2 1/2}^{T' M_{T'}} \sum_{M_S M_S'}C_{S M_S \lambda \mu}^{J M} C_{S' M_S' \lambda ' \mu '}^{J M}
\langle (1/2\ 1/2) T M_T |\langle (s_a s_b) S M_S (l_a l_b) \lambda \mu | V(1-\hat{P}^r \hat{P}^\sigma \hat{P}^\tau) \\| (s_c s_d) S' M_S ' (l_c l_d) \lambda ' \mu '\rangle | (1/2\ 1/2) T' M_{T'} \rangle.
\label{natSep1}
\end{multline}
In the second step one needs to substitute $V$ with generic separable form, similar to Eq. \eqref{pair_int} but now without any kind of projection,
\begin{equation}
V(\vec{r}_1, \vec{r}_2, \vec{r}_1', \vec{r}_2')~=~-G_0 \delta(\vec{R}_1-\vec{R}_2) G(r) G(r')
\end{equation}
into Eq. \eqref{natSep1} and transform the laboratory coordinates $\vec{r}_1$ and $\vec{r}_2$ ($\vec{r}_1'$ and $\vec{r}_2'$) in \textit{bra} (\textit{ket}) into the center of mass $\vec{R}$ ($\vec{R}'$) and relative coordinates $\vec{r}$ ($\vec{r}'$), i.e. the so-called Talmi-Moschinsky transformation. Therefore we need to evaluate
\begin{align}
 \sum_{T M_T} \sum_{T' M_{T'}} &C_{1/2 -1/2 1/2 1/2}^{T M_T}C_{1/2 -1/2 1/2 1/2}^{T' M_{T'}}
\langle (1/2\ 1/2) T M_T | \langle (1/2\ 1/2) S M_S (l_a l_b)\lambda \mu | \nonumber \\ &V(1-\hat{P}^r \hat{P}^\sigma \hat{P}^\tau) |  (1/2\ 1/2) S' M_{S'} (l_c l_d)\lambda' \mu' \rangle | (1/2\ 1/2) T' M_{T'} \rangle~=~
\nonumber
\\~=&-4\pi G_0 \delta_{SS'} \delta_{M_S M_{S'}} \delta_{\lambda L} \delta_{\mu m_L} \delta_{\lambda' L} \delta_{\mu' m_{L'}} 
\sum_{NLnn'} I_n I_{n'} M_{n_a l_a n_b l_b}^{N L n 0} M_{n_c l_c n_d l_d}^{N L n' 0} 
\nonumber
\\ &\sum_{T}\frac{1}{2}\left(1+(-1)^{S'+T+1}\right)
\label{natSep2}
\end{align}
Constraining ourself to the proton-neutron case only, note that the only nonvanishing cases of the isospin coupling are $T=0$ and $M_T=0$ or $T=1$ and $M_T=0$, which both lead to factor $1/2$ in front of the round brackets in Eq. \eqref{natSep1}. Therefore, substituting back Eq. \eqref{natSep2} into Eq. \eqref{natSep1} after some mathematical manipulations we obtain
\begin{align}
V_{abcd}^{J M}~=&~-G_0\hat{j}_a \hat{j}_b \hat{j}_c \hat{j}_d \sum_{L S} \sum_{T}\frac{1}{2}\left(1+(-1)^{S'+T+1}\right) \tilde{f}(S,T) \hat{S}^2 \hat{L}^2\left\{\begin{array}{clcr}
l_b & 1/2 & j_b\\
l_a & 1/2 & j_a\\
L & S & J\end{array}\right\}
\left\{\begin{array}{clcr}
l_d & 1/2 & j_d\\
l_c & 1/2 & j_c\\
L & S & J\end{array}\right\}
\nonumber\\
&~~~\sum_{n n'} \tilde{I}_n \tilde{I}_n' M_{n_a l_a n_b l_b}^{N L n 0} M_{n_c l_c n_d l_d}^{N L n' 0}.
\label{total_pairing2}
\end{align}
However, we also add multiplication with function $\tilde{f}(S,T)$ which should take into account somewhat smaller or enhanced effect of the pairing in $T ~=~ 0$ case,
\begin{equation}
  \tilde{f}(S,T)~=~
  			\begin{cases}
               1,~{\text{for} ~S=0~,~T=1}\\
               V_{0pp},~{\text{for}~S=1~,~T=0}\\
               0,~{\text{the rest}}
            \end{cases}
\end{equation}
while $\tilde{I}_n~=~\sqrt{4\pi}I_n$ is spatial integral of Gaussian $G(r)$ derived analytically in the next section of Appendix.

\section{Analytical solution of radial part}
Radial part of the eigenfunction of spherical 3D harmonic oscillator is given by
\begin{equation}
R_{n l}(r, b_0)~=~b_0^{-3/2} R_{n l}(\xi^2) ~=~ b_0^{-3/2} N_{n l} \xi^l L_{n}^{l+1/2}(\xi^2) e^{-\xi^2/2},  
\end{equation}
where $b_0$ represents oscillatory length, while $n$ corresponds to the nodes number (in this notation we don't take into account 0 and $\infty$). Possible values are $n = 0, 1, 2 ...$, and $\xi = r/b_0$. Normalization factor $N_{nl}$ is given by
\begin{equation}
N_{n l}~=~\left( \frac{2 n!}{\Gamma(l+n+3/2)} \right)^{1/2}  
\label{sho_norm}
\end{equation}
In the case of half-number arguments gamma function is given by
\begin{equation}
\Gamma\left( \frac{1}{2} + n\right)~=~ \frac{1\cdot 3\cdot 5  \cdots (2n-1)}{2^n}\sqrt{\pi}~=~\frac{(2n-1)!!}{2^n}\sqrt{\pi},
\label{gamma}
\end{equation}
We are interested in the analytical solution of the following integral
\begin{equation}
I_n ~=~ \int R_{nl}(r) G(r) r^2 dr
\label{anInt}
\end{equation}
By inserting G(r) from Eq. \eqref{Gaussian} in Eq. \eqref{anInt} we obtain
\begin{equation}
I_n~=~\frac{b_0^{3/2}N_{n 0}}{(4\pi a^2)^{3/2}}\int_0^\infty \exp\left[-\frac{\xi^2}{2}\left(\frac{b_0^2}{a^2}+1\right)\right] L_n^{1/2}(\xi^2) \xi^2 d\xi.
\label{startIn} 
\end{equation}
After substitution 
\begin{equation}
\alpha^2~=~\frac{a^2}{b_0^2},~~ \xi ^2 =\eta\rightarrow \xi d\xi = \frac{d\eta}{2}, 
\end{equation}
Eq. \eqref{startIn} can be rewritten as
\begin{equation}
I_n~=~\frac{b_0^{-3/2} N_{n0}}{(4\pi\alpha^2)^{3/2}} \frac{1}{2}\int_0^\infty \exp\left[-\eta + \frac{\eta}{2}\left(1 - \frac{1}{\alpha^2} \right)\right] L_n^{(1/2)}(\eta)\sqrt{\eta}d\eta.
\label{startIn2}
\end{equation}
Generating function for Laguerre polynomials with $\alpha~=~1/2$ ($l ~=~ 0$) is given by expression \cite{Abram}
\begin{equation}
\frac{1}{(1-z)^{3/2}}\exp\left(\frac{xz}{z-1}\right)~=~ \sum_{n=0}^\infty L_n^{1/2}(x) z^n.
\label{genfun}
\end{equation}
By using substitution
\begin{equation}
\frac{1}{2}\left(1-\frac{1}{\alpha^2}\right)~=~ \frac{z}{z-1} ~\rightarrow~
z ~=~ \frac{1-{\alpha^2}}{1+{\alpha^2}}
\end{equation}
and Eq. \eqref{genfun}, we can rewrite Eq. \eqref{startIn2} in the following form,
\begin{equation}
I_n~=~\frac{b_0^{-3/2} N_{n0}}{(4\pi\alpha^2)^{3/2}} \frac{1}{2} (1-z)^{3/2}\sum_{m=0}^\infty z^m\int_0^\infty d\eta \eta^{1/2} e^{-\eta} L_n^{(1/2)}(\eta) L_m^{(1/2)}(\eta).
\label{Pbefore}
\end{equation}
The integral in Eq.\eqref{Pbefore} is just orthogonality condition for Laguerre's polynomials \cite{Abram},
\begin{equation}
\int_0^\infty x^\alpha e^{-\alpha} L_n^{\alpha}(x) L_m^{\alpha}(x) dx~=~ \frac{\Gamma(n+\alpha+1)}{n!}\delta_{nm}.
\end{equation}
Therefore, Eq. \eqref{Pbefore} reduces to
\begin{equation}
I_n~=~~\frac{1}{2}\frac{b_0^{-3/2} N_{n0}}{(4\pi\alpha^2)^{3/2}} (1-z)^{3/2}\sum_{m=0}^\infty z^m \frac{\Gamma(m+3/2)}{m!} \delta_{nm}.
\end{equation}
Using Eq. \eqref{gamma}, inserting normalization factor (Eq. \eqref{sho_norm}) and expressing everything in terms of $\alpha$ and $n$, after some mathematical manipulations, it can further be simplified to
\begin{equation}
I_n~=~\frac{1}{2^{5/2} \pi^{5/4} b_0^{3/2}} \frac{(1-\alpha^2)^n}{(1+\alpha^2)^{n+3/2}} \frac{\sqrt{(2n+1)!}}{2^n n!}.
\end{equation}

%%%%%%%%%%%%%%%%%%%%%%%%%%%%%%%%%
% bibliography
%
\bibliographystyle{apsrev4-1}
%\bibliography{bibliography}
\bibliography{pcpnqrpa_NP.bbl}

\end{document}